\documentclass[12pt,preprint,referee]{aastex}

\RequirePackage{lineno}
   \usepackage{color}

\usepackage{gensymb}
 \usepackage{lscape}

\title{
	\Large
	Search for transitions between states in
	redbacks and black widows using seven years of \emph{Fermi}-LAT observations
}

\author{Diego F. Torres\altaffilmark{1,2}, Long Ji\altaffilmark{3,4}, Jian Li\altaffilmark{1}, Alessandro Papitto\altaffilmark{5}\\
Nanda Rea\altaffilmark{1,6}, Emma de O\~na Wilhelmi\altaffilmark{1}, Shu Zhang\altaffilmark{3} }
\altaffiltext{1}{Institute of Space Sciences (IEEC-CSIC), Campus UAB, Carrer de Magrans s/n, 08193 Barcelona, Spain}
\altaffiltext{2}{Instituci\'o Catalana de Recerca i Estudis Avan\c{c}ats (ICREA), E-08010 Barcelona, Spain}
\altaffiltext{3}{Key Laboratory for Particle Astrophysics, Institute of High Energy Physics, Chinese Academy of Sciences, 19B Yuquan Road,
Beijing 100049, China}
\altaffiltext{4}{Institut f\"ur Astronomie und Astrophysik, Kepler Center for Astro and Particle Physics, Sand 1, D-72076 T\"ubingen, Germany}
\altaffiltext{5}{INAF-Osservatorio Astronomico di Roma, via di Frascati 33, I-00040 Monte Porzio Catone, Roma, Italy}
\altaffiltext{6}{Anton Pannekoek Institute, University of Amsterdam, Postbus 94249, NL-1090-GE Amsterdam, The Netherlands}
\begin{abstract}
Considering {about seven} years of \emph{Fermi}-{Large Area Telescope (LAT) }data, we present a
systematic search for variability possibly related to transitions between states in redbacks and black
widow systems.
{Transitions} are characterized by sudden and significant changes in the gamma-ray flux that persist {on} a timescale much larger than the orbital
period.
This phenomenology was already detected in the case of two redback systems, PSR J1023+0038 and PSR J1227$-$4853, for which we present  here a dedicated
study. %
We show the existence of only one transition for each of these systems
{over the past} seven years. We
determine their spectra, establishing high-energy
cutoffs
at a few GeV for the high gamma-ray state of PSR J1023+0038 and for both states
of PSR J1227$-$4853.
The surveying capability of {the} \emph{Fermi}-LAT allows studying {whether similar phenomenology has occurred} in other sources.
{Although we have not} found any hint for a state transition for most
of the studied pulsars,
we note two black-widow systems, PSR J2234+0944 and PSR J1446$-$4701, {whose apparent variability is} reminiscent of the transitions in PSR J1023+0038 and PSR J1227$-$4853.
For the other systems we set limits on potential transitions in their measured gamma-ray light curves.

\end{abstract}

\keywords{gamma-ray: observations, pulsars: individual (PSR J1023+0038, PSR J1227$-$4853)}

\begin{document}
\maketitle
\section{Introduction}

{In the framework of the recycling pulsar model,
the rapid spinning of old millisecond pulsars {is} the outcome of
accretion of mass transferred by a low-mass late-type companion star onto a neutron star (NS) via an accretion disk (Alpar et al. 1982).}
After a Gyr-long mass accretion phase during
which the binary system shines as a bright low mass X-ray binary
(NS-LMXB), the mass transfer rate declines and allows the activation
of a radio/gamma-ray millisecond pulsar (MSP)
powered by the rotation of its magnetic field.

The tight link existing
between radio MSPs and NS-LMXBs has been only recently demonstrated by the
discovery of three transitional millisecond pulsars (PSR J1023+0038,
Archibald et al. 2009; {PSR J1824$-$2452I} in the
globular cluster M28, Papitto et al. 2013; and XSS J12270$-$4859 / PSR J1227$-$4853, de Martino et al. 2010, 2013, 2015, Bassa et al. 2014).

These sources have been observed to switch between
accretion and rotation-powered emission on timescales possibly shorter
than a couple of weeks. Thus, such state transitions may take
place on timescales compatible with those of the variations of the
mass accretion rate onto the NS.

Under this hypothesis (see e.g., Bogdanov 2015 and other references above), at high mass inflow rates, the radio
pulsar is shut-off and the system is bright in X-rays ($L_X > 10^{36}$
erg s$^{-1}$). At low mass inflow rates, the magnetosphere probably expands up to the
light cylinder, activating the radio pulsar; the disk disappears and
the system is instead {quiet} in X-rays ($L_X \sim 10^{32}$ erg/s).

In
addition to the accreting and the radio pulsar
states, the transitional pulsars known so far  have been
observed to enter into a sub-luminous disk state with
$L_X \sim 10^{33}$ erg/s.
During this state, both J1023+0038 (Archibald et
al. 2015; {in the text below we will omit the ``PSR'' or ``3FGL'' prefix of the sources for simplicity}) and J1227$-$4853 (Papitto et al. 2015) have 6--8\%
of the X-ray flux pulsed. Most likely, then, a part of
the disk material is accreted onto the NS surface.
In addition,
these sub-luminous
states are accompanied by a sizeable gamma-ray flux and a {flat radio}
spectrum that are typical signatures of jets  in accreting compact
objects, suggesting that large mass outflows could be launched by
these pulsars.
These features prompted theoretical models based on the propeller mechanism
 (Papitto, Torres, \& Li 2014; Papitto \& Torres 2015, Campana et. al. 2016).

Currently, the timescales of the transitions are only loosely
constrained, but are of primary importance
because they reflect the timescale of disk formation and the
transition into a jet-dominated outflow or of the disk evaporation.
Candidate transitional pulsars are identified among binary radio
millisecond pulsars whose donor {stars are} currently losing mass, as
indicated by the irregular eclipses of the radio pulsed emission
caused by absorption and scattering by the matter ejected from the
system.
{They are tight binaries (P$_{orb} <$ 1 day), and are dubbed black widows (BWs, M$_{companion} <<$ 0.1 M$_{sun}$; Fruchter et al. 1988) or redbacks (RBs, M$_{companion} \sim $ 0.2 -- 0.4 M$_{sun}$; D'Amico et al. 2001), depending {on} the mass of the companion.}
Tens of systems of this kind were discovered by radio surveys in our Galaxy (see the ATNF Pulsar Catalog\footnote{\url{http://www.atnf.csiro.au/people/pulsar/psrcat/}}, {Manchester et al. 2015}).
They possess {a} similar spin distribution, {which is} intermediate between the faster {accretion}-powered millisecond pulsars and the slower non-eclipsing rotation-powered millisecond pulsars, and are {evolutionarily }linked (Chen et al. 2013, Papitto et al. 2014).

{Here we analyze \emph{Fermi}-LAT data to systematically search
for transitions, }not only in the known transitional pulsars,
but in a large sample of {RBs} and {BWs}.

\section{Observations and data analysis}

The \emph{Fermi}-LAT data included in this report {cover} nearly {seven} years, from 2008 August 4 (MJD 54682) to 2015 June 1 (MJD 57198).
The analysis of \emph{Fermi}-LAT data was performed using the \emph{Fermi} Science Tools 10-00-05 release\footnote{\url{http://fermi.gsfc.nasa.gov/ssc/}}.
Events from the ``Pass 8'' event class were selected.
The ``Pass 8 R2 v6 Source'' instrument response functions (IRFs)
were used in the analysis.
We have considered all gamma-ray photons within {the energy range} 0.1--300~GeV {for circular regions} of
interest (ROI) of 10$\degree$ radius around each of the RBs \& BWs.
Additionally, to reject contaminating gamma rays from the Earth's limb,
we have only selected events with zenith angle $<$ 90$\degree$.
The systematic errors have been estimated by repeating the analysis using modified IRFs that bracket the effective area (Ackermann et al. 2012), and artificially changing the normalization of the Galactic diffuse {emission} model  by $\pm$6\% (Abdo et al. 2013).
However, note that {changes of measured fluxes evaluated} with and without systematic uncertainties considered are far lower than the flux jumps we are {seeking} here, i.e., usually {few percent compared to a} jump in flux by a factor of several.

The gamma-ray fluxes presented in this work were calculated by performing a binned
maximum likelihood fit using the Science Tool \emph{gtlike}.
The spectral-spatial model constructed to perform the likelihood analysis includes Galactic and isotropic diffuse emission
components {(``gll\_iem\_v06.fits", Acero et al. 2016, and ``iso\_P8R2\_SOURCE\_V6\_v06.txt", respectively\footnote{\url{http://fermi.gsfc.nasa.gov/ssc/data/access/lat/BackgroundModels.html}})}, as well as known gamma-ray sources within 15$\degree$ of the source analyzed, as included in the third {\it Fermi} Source Catalog
(Acero et al. 2015, 3FGL hereafter).
The spectral parameters and the positions of all gamma-ray sources were fixed to the catalog values, except for those within 3$\degree$ from the considered targets.
For these latter
sources, the spectral parameters were left free.
In the cases of previously known associations between gamma-ray sources and RBs or BWs (see Table~1), we have
adopted the spectral shape reported in the 3FGL.
If no association was {made earlier}, their spectra were modelled with a simple power law.
All spectral parameters of RBs and BWs were allowed to vary.

The test statistic (TS, {Mattox et al. 1996}) was employed to evaluate the significance of the gamma-ray fluxes from the sources.
The TS is defined by TS=$-2 \ln (L_{max, 0}/L_{max, 1}) $, where $L_{max, 0}$ is the maximum likelihood value for a model without an additional
source (the ``null hypothesis") and $L_{max, 1}$ is the maximum likelihood value for a model with the additional source at a specified location.
A larger TS indicates that the null hypothesis is not preferred. {The TS is distributed as $\chi^{2}$ so that a gamma-ray excess at the tested position can be deemed significant if TS $>$ 25.}

\section{Search for state transitions in long-term light curves}

We have carried out the analysis of the 12 confirmed RBs and the 18 confirmed BWs (Table~\ref{table1}).
{The transitional millisecond pulsar PSR J1824$-$2452I is not included in the analysis,
since it is located in the globular cluster M28, which is a bright gamma-ray source (Abdo et al. 2010a).}
To search for possible state transitions in RBs and BWs, we  produced long-term light curves for all systems in Table~\ref{table1}.
{The average TS values, spectral parameters, and average fluxes} along the whole observation period for all systems in our search
are shown in Table~\ref{table1}.

RBs and BWs usually host gamma-ray pulsars.
{Eleven} out of 12 RBs, and 15 out of the 18 BWs are significantly detected in the 0.1--300~GeV band.
In the second \emph{Fermi} Large Area Telescope Catalog of Gamma-Ray Pulsars (Abdo et al. 2013, 2PC {hereafter}),
pulsations in gamma rays {were} identified in {one out of four} RBs {known at that time}, and 10 out of 16 BWs {known at that time}.
With more \emph{Fermi}-LAT data accumulated ({seven} years against the {three} years considered in the 2PC), the use of a more advanced event-level analysis (Pass 8 against Pass 7), and a better modelling of the {gamma-ray sky} (3FGL against 2FGL), more gamma-ray pulsars were found in known RBs and BWs by the \emph{Fermi}-LAT collaboration\footnote{\url{https://confluence.slac.stanford.edu/display/GLAMCOG/Public+List+of+LAT-Detected+Gamma-Ray+Pulsars}, noted as ``list'' in {Table~\ref{table1}}.}
(Table~\ref{table1}).
Currently, {seven} out of 12 RBs, and 15 out of 18 BWs are already known to pulse in gamma rays, which clearly contribute to the detected gamma-ray emission.
The current {non-detection} of gamma-ray pulsars {among} the rest of the RBs and BWs may be a result of limited gamma-ray statistics, not well known pulsar ephemerides and/or binary parameters, or an unfavorable beaming direction of the pulsar gamma-ray emission with respect to our line of sight ({Guillemot \& Tauris 2014}).

Using {distances} from the radio dispersion measure when available\footnote{http://www.atnf.csiro.au/research/pulsar/psrcat/}({Cordes \& Lazio 2002}), we converted the gamma-ray {fluxes} into {corresponding {luminosities} ($L_{\gamma}$), which are }also listed in Table~\ref{table1}.
The gamma-ray {luminosities {(L$_{\gamma}=4\pi D^{2} F$, $D$ and $F$ are} the distance and flux in Table \ref{table1})} of RBs and BWs range from $\sim$10$^{33}$ erg s$^{-1}$ to $\sim$10$^{35}$ erg s$^{-1}$.

During the state transitions of the few transitional millisecond pulsars known  (i.e., in the {cases} of J1023+0038, Stappers et al. 2014; and J1227$-$4853, Johnson et al. 2015; both being RB systems) their gamma-ray flux was observed to vary by a factor of 2 to 5.

\subsection{A fixed time binning and analysis of the known transitional pulsars}

We {initially} used a time bin of 60 days to construct the light curves (Fig.~\ref{LC}).
This was the timescale earlier used to analyze and discover the known transitional pulsars (see, e.g., Stappers et al. 2014, Bogdanov \& Halpern 2015, also Bogdanov 2016).
{It is also a sensible time bin selection for the average fluxes presented here;} if the time bin is significantly {shorter} than this timescale, there would be too few counts per bin in many cases; if {it is} significantly
larger than this timescale, there would be {fewer bins to search for} the {light curve} evolution and transitions of duration of up to several months could be missed.
Below
we also explore other timescales chosen on a case-by-case basis.
{Flux upper limits (95\% confidence level)} were calculated using Helene's method (Helene 1983), assuming the photon index in Table~\ref{table1} if the TS value of a time bin is below 12 ($\sim$3.5$\sigma$, dotted green line in Fig.~\ref{LC}).

Significant flux variations are {found} in this analysis only in the {cases} of J1023+0038 and J1227$-$4853, at the already known state transitions times (see Fig.~\ref{LC}, where the dotted vertical line indicates the time at which they happen).
These results are in agreement with {those of} Stappers et al. (2014) for J1023+0038 and Xing \& Wang (2015) for J1227$-$4853.
Note that the transitions in  J1023+0038 and J1227$-$4853 occur in opposite directions; i.e., while
for  J1023+0038 the transition is from a {radio pulsar state (low gamma-ray state) to a sub-luminous disk state (high gamma-ray state)}, it is the opposite for J1227$-$4853.
{Assuming the same physical mechanism is at work in both sources, the back and forth nature of the swinging phenomenon is emphasized by
this fact.}
No other transition at earlier or later times is discovered in the additional dataset analyzed here.

We modelled J1023+0038 and J1227$-$4853 while being in {radio pulsar and sub-luminous disk states} using a power-law function with and without an exponential cutoff.
A spectral cutoff at E$_{cut}$=3.7 $\pm$1.3 $\pm$ 0.9 GeV (as usual the first error comes from statistics and the second from systematics)
was detected  for J1023+0038 in the {sub-luminous disk state}, with a spectral index of 2.0 $\pm$ 0.1 $\pm$ 0.1.
{The likelihood ratio test indicates} a $\Delta$TS of 18.9, which indicates that the significance of the spectral cutoff is $\sim$4.3$\sigma$.
No spectral cutoff is detected for J1023+0038 during its {radio pulsar state},
which is consistent with the reports by Takata et al. (2014) and Tam et al. (2010).
The SEDs of J1023+0038 during {radio pulsar state} and {sub-luminous disk state} are shown in Fig. \ref{SED}.

Spectral cutoffs are hinted at for J1227$-$4853 during both periods, before and after the gamma-ray transition.
Before the transition, in the {sub-luminous disk state} of the source, {\ a spectral cutoff at E$_{cut}$=10.8 $\pm$ 3.7$\pm$ 5.6
GeV, with a power-law having a spectral index of 2.3 $\pm$ 0.1 $\pm$ 0.1, yields a $\Delta$TS of 13.9 (which implies it is a better fit than a simple power-law
at $\sim$3.7$\sigma$).}
After the transition, in the {radio pulsar state} of the source, the existence of a cutoff provides a $\Delta$TS of 11.4 ($\sim$3.4$\sigma$), so that J1227$-$4853 is better described by a power law with cutoff at E$_{cut}$=5.3 $\pm$ 2.5 $\pm$ 2.3 {GeV} and spectral index of 2.0 $\pm$ 0.1 $\pm$ 0.3.
The cutoff energies and spectral indices before and after the transition are compatible within their corresponding 1$\sigma$ errors, and are consistent with the values {reported} by Johnson et al. (2015) and Xing \& Wang (2015).

\subsection{A flux-motivated definition of the time binning}

Using the fitted model from the {approximately  seven} years
binned maximum likelihood analysis and the tool \emph{gtobssim},
we produced simulated {\it Fermi}-LAT observational data for each of the sources of interest.
Simulated data were produced for different exposure times, in a similar region of interest to the real data, and were analyzed following the standard steps described in Section 2.
{The simulated TS values of RBs and BWs were obtained for the different exposure times assigned in the simulation, and were fitted by a linear function.}
With the fitted linear function, we estimated the observation time needed for a {TS=25-detection with the \emph{Fermi}-LAT} at
90\% confidence level
for the average level of flux given in Table~1.
This estimation gives a natural time binning for each source that is adequate {for} the corresponding level of flux:
under the assumption of no variability, the extent of each time bin should be enough {to result in} a detection of the source with a TS value around 25.
With the time binning so defined for each source, we produced the corresponding long-term light curves.

To each of these light curves, we fitted a constant and computed the $\chi^2$, as described in Abdo et al. (2010b).
Table~2 gives for each source the results of this fitting, and the time bin used for each source.
In particular,  the $P$-value ({{probability} of the flux being constant}) and the corresponding $\sigma$ {indicate} the significance of
flux variability.
We see that the obtained light curves are all compatible with no persistent jump in flux (no transition) except for the known transitional pulsars
and two other cases.

The case of the transitions in J1023+0038 and J1227$-$4853 is shown in Fig. \ref{odd} (upper panels). For the former, {applying the simulation-determined binning of 9 days results in a number of upper limits before the transition and a number of significant detections just after; the transition is obvious.}
The smaller variation of flux between the high and the low gamma-ray state of
J1227$-$4853 allows for significant detections before and after the transition. The {bin width} resulting from the simulations is in this case
larger (53 days).
Table~2 shows that for J1227$-$4853, {the hypothesis of constant flux is rejected with $\sim$10 $\sigma$ confidence}.
In the low state, the flux evolution is compatible with a constant.
However, in the high gamma-ray state of J1227$-$4853, a constant flux is put in question (ruled out with
a significance of 4.7$\sigma$), likely indicating {the action of shorter-timescale phenomena}.

The variability appearing in J1446$-$4701 and J2234+0944  seems reminiscent of the phenomenology shown by
J1023+0038 and J1227$-$4853, despite the caveat (especially for J1446$-$4701) of their {lower fluxes}.
Their light curves are shown in Fig. \ref{odd} (lower panels). A constant line fitting is ruled out at 4.2$\sigma$ and 5.5$\sigma$, respectively.

The case of J1446$-$4701 is
difficult to assess due to its low flux. This is translated into the large time binning (506 days) needed to achieve an individually significant detection at its average {flux level}.
{With the difference between the putative {\it low} and {\it high} gamma-ray states (about a factor of 2), a firm conclusion cannot be drawn because of the {lack} of consecutive data points at the {\it high state}.}

{The phenomenology of J2234+0944 in gamma rays is clearer, since {the greater average flux} allows for a detection in a time bin  {of 106 days}, a factor of $\sim 5$ shorter than that needed for J1446$-$4701. Given that this timescale is larger than 60 days,
the possible variation of J2234+0944 was not hinted at in the fixed time bin study of the earlier section:
The shorter timescale leads to larger error bars, which hide the possible flux variation}.
The J2234+0944 {light curve} shown in Fig. \ref{odd} {resembles} that of J1227$-$4853 (see second panel of Fig. 2).
With {constant flux being excluded} (at 5.5$\sigma$), two horizontal lines separated around {MJD 55500} ({2010 October 31}) provide correspondingly good fits: the source seems to have jumped to a higher level of gamma-ray flux (a factor of $\sim$2 higher) then.
{Interestingly, we note that in the two periods separated by MJD 55500, gamma-ray pulsations from J2234+0944 are significantly detected at significance greater than 8$\sigma$ level, {respectively}. }
{We have produced orbitally-folded light curves of J2234+0944, with an orbital period of 0.42 days (Ray et al. 2012) using all data, as well as the data before and after MJD 55500. No significant orbital modulation is detected before or after MJD 55500.}

In order to assess how the variability shown in Fig. \ref{odd} could indicate a state transition we analyzed existing X-ray observations.
{In the period covered by \emph{Fermi}-LAT data, there were two relatively deep X-ray observations of J1446$-$4701, done with \textit{XMM-Newton} and \textit{Swift}/XRT.
The \textit{XMM-Newton} observation was carried out on 2012 {August}({MJD 56140}) and has 62 ks of exposure (Arumugasamy et al. 2015). The \textit{Swift}/XRT observation was carried out in 2012 {January}({MJD 55932}) and has 9.8 ks of exposure.
However, both {observations were made} in the putative {\it low} gamma-ray state of the source.
J1446$-$4701 was detected by \textit{XMM-Newton} with a very low flux level of about 9$\times 10^{-14}$ erg cm$^{-2}$  s$^{-1}$ in 0.5--10 {keV} (Arumugasamy et al. 2015).
In the \textit{Swift}/XRT observation J1446$-$4701 was not detected at all, yielding a compatible 95\% flux upper limit in 0.5--10 keV band of about 1.2$\times 10^{-13}$ erg cm$^{-2}$  s$^{-1}$, assuming a power-law spectrum with index of 2.}

{Two observations of J2234+0944 with \textit{Swift}/XRT were carried out} in October 2013({MJD 56583}) and April 2016({MJD 57497}); both {lasted} less than 6 ks.
J2234+0944 was not detected in either of the observations, leading to 95\% flux upper limits in 0.3--10 keV band of about 3$\times 10^{-13}$ erg cm$^{-2}$  s$^{-1}${, also assuming a power-law spectrum with index of 2.}

Even in the rotationally-powered state, the flux in the 0.5--10 keV range of J1227$-$4853
was determined to be larger than these measurements:
7.0(5)$\times 10^{-13}$ erg cm$^{-2}$  s$^{-1}$ (Papitto et al. 2015). The flux from J1227$-$4853 in the same {energy} band in the high gamma-ray state  is larger by more than one order of magnitude
(Papitto et al. 2015).
The {best estimates of the distances} based on the dispersion measure are 1 kpc for J2234+0944 (Ray et al. 2012) and 1.4 kpc for J1227$-$4853
(Roy et al. 2015). In the disk state, J1227$-$4853 emits {an} 0.5--10 keV unabsorbed flux of $\sim 2 \times 10^{-11}$ erg cm$^{-2}$  s$^{-1}$ (Papitto et al. 2015). Therefore, the upper limit on the luminosity of J2234+0944 is
$\sim 5 \times 10^{-3}$ times smaller than the luminosity observed from J1227$-$4853.
Such a low X-ray flux argues against the presence of an accretion disk for a source this close. A larger distance, however, would obviously {help in making} the fluxes more compatible.
For instance, at a distance of $\sim$3 kpc, the difference in the luminosities would be around one order of magnitude.
Unless we are witnessing a transition at a much lower level of X-ray flux,
a possible transition similar to that in J1227$-$4853 is put in doubt.

We note that the above discussion on  J1446$-$4701 and J2234+0944
does not mean that there is no intrinsic variability in other sources.
Rather, any variability is within the errors of the fitting (i.e., within the normal evolution of the fluxes, without the appearance of a persistent jump).
In particular, all sources can have transitions at shorter timescales than {\it Fermi}-LAT is able to test, resulting in an intrinsic flux dispersion in the time binning explored, without the appearance of an ordered transition as shown by the transitional pulsars known.
In order to consider what kind of transitions {\it can} be detected for other sources, Table~2 shows the
{\it jump factor}.
This factor indicates the level of flux
that would deviate from its mean by $3\sigma$, assuming that the increased flux has the same uncertainty as the measured one and adding in quadrature the error of the fitted constant.
Subsequently,
Table~2 gives the number of data points that would be needed in the {light curve} at this increased flux level  to be able to claim a 3$\sigma$ overall variability for the {light curve} as a whole.
Put otherwise, if we fit
with a single horizontal line the whole set of points, this fitting would be ruled out with a significance larger than 3$\sigma$.
We can thus quantitatively conclude that none of these transitions (nor others more significant than these) have occurred.

\subsection{Conclusions}

We {temporally} enlarged the analysis of both J1023+0038 and J1227$-$4853, the known transitional pulsars, by considering nearly {seven} years of \emph{Fermi}-LAT data.
Our results on the {light curves} of these systems confirmed previous reports.
Only one transition is detected for each.
We found that they transitioned from a low to a high state, and from a high to a low state, respectively.
In addition, we determined their spectra, and confirmed the existence of high-energy cutoffs at a few GeV with the significance above 3$\sigma$ for the high gamma-ray state of J1023+0038 and for both states of J1227$-$4853.

We searched for state transitions in all known RBs and BWs by analyzing their long term light curves in different time binnings.
Our analysis included a fixed 60-day time {binning}, used {for} detection of the already known transitional pulsars mentioned above.
We have also performed simulations for each source in order to determine, assuming their average level of flux, the minimum integration time needed for a {\it Fermi}-LAT detection at a TS=25 level.
This is a flux-motivated, source-by-source-determined binning, and {we} have used it to study the light curves as well.
By analyzing the  light curves we were able to determine whether a transition has happened and if not, what are the features of the transitions that can be ruled out.

For most of the pulsars, {we have not found any hint} for a state transition in our search.
In the light of negative results, trying to {infer conclusions} regarding e.g., rate of transitions, seems daunting.
Transitions are inextricably linked to the local
scenario, for instance, to the variations in mass accretion rate.
A negative result cannot be directly used to imply that all {RBs} and {BWs} other than the swinging ones,
have actually finished any swinging phase, and are all in a final -fully recycled- state.
Future surveying may prove the opposite, and when this swinging {will} happen, if it does, can simply not be predicted.

We found two particularly interesting cases in our search.
J2234+0944 and J1446$-$4701 are, in contrast with the known transitional pulsars, BW systems. Both of these sources have very low companion masses.
Both were discovered at Parkes as part of a radio search program for pulsars in coincidence with unidentified {\it Fermi}-LAT sources (see Ray et al. 2012).
{The radio detection of J2234+0944 was before the possible transition at MJD 55500.}

J2234+0944 has a period of 3.63 ms and is part of a system with a companion of at least 0.015 M$_\odot$, in an orbit of 0.42 days.
J1446$-$4701 is in a system with a companion of at least 0.019 M$_\odot$, in an orbit of 0.27 days (Keith et al. 2012). The orbits are almost circular, which is
consistent with the model in which the spin-up of the pulsar is associated with Roche lobe overflow from a nearby companion.
Both orbital periods are much smaller than the timescale for the variability we have {found. Thus} the latter can hardly have an orbital origin.
But are these indeed {\it state transitions} similar to those found in
J1023+0038 and J1227$-$4853?

The  variability of J1446$-$4701 is not conclusive, although {a possible back-and-forth flux jumps during the years spanned by \emph{Fermi} observations}
is compatible with the data. Inconclusiveness arises from the fact of it being a very dim source in comparison to the known transitional pulsars (see Table~1), and from the (related) {lack of} a sufficient number of points
in each of the putative states. {We recommend further monitoring of this source} in {gamma rays} and other frequencies.
The variability of J2234+0944 is clearer, and a flux jump seems to have happened (see Table~2 and Fig. \ref{odd}).
Its brightness in gamma rays allows for a clear distinction of two states that can be deemed similar to those in J1227$-$4853, with an apparent {transition} from lower to higher gamma-ray fluxes.
However, the low level of fluxes found in existing X-ray observations cast doubts {that} we are witnessing the same phenomenology.
Future X-ray observations will tell whether this pulsar has a short timescale phenomenology as that found for J1023+0038 and J1227$-$4853,
yet at a significantly lower level of flux.
If not, we may be
witnessing a gamma-ray state transition produced at the intra-binary shock and/or with {a} dim (if any) counterpart at lower frequencies.
The latter would not be impossible within the propeller {model} used to investigate J1023+0038 and J1227$-$4853. If the propeller is {strong enough} to preclude any matter {from} reaching the surface
and the disk component is significantly {dimmer in X-rays} in comparison with redback systems, it is in fact expected that the X-ray emission would be undetectable, smaller {by even more} than several orders of magnitude in comparison with that of J1023+0038 and J1227$-$4853 (see {Fig. 1} in Papitto \& Torres 2015).
A proper model, together with deep X-ray observations would help {test} this setting.
Alternatively, we can also entertain the possibility that
the pulsar magnetosphere could globally vary  ({see, e.g.,} the study by
Ng et al. 2016, {even though} it is a very different system). If this is the case, the variation in the two states
would be explained by a closer-to-the-pulsar phenomenology, and could just be interpreted as being different
{outer gap-generated} emission, as if the pulsar would be isolated.
This would naturally encompass the fact that gamma-ray pulsations are found both before and after the flux jump.

\begin{landscape}
	\begin{table}[hbt!]
		\scriptsize
		\centering
		\caption{RBs and BWs included in this report. Columns denote the source name, the 3FGL association, TS value, index, flux level, Right Ascension and Declination {(J2000)}, whether it is detected in the 2PC/\emph{Fermi}-LAT public pulsar catalog {as of February 2016}, luminosity, and distance (from the ATNF catalog). $\dag$ and $\ddag$ indicate the data period before and after the state transition for known transitional millisecond pulsars. The first/second error is statistical/{systematic}.}
				\vspace{0.1cm}
		\label{table1}
		\begin{tabular}{llcccccccc}
\hline
RB name               &   3FGL source    &  TS        & Index            & Flux                                            & $RA$     &   $Dec$   & 2PC/list                &   L$_{\gamma}$        &  D     \\
                                   &                  &            &                  & (10$^{-11}$ erg cm$^{-2}$  s$^{-1}$)                &   deg   &   deg    &            &   ($10^{33}$ ergs/s) &   (kpc)        \\		
		\hline
3FGL J0523.3$-$2528        &  J0523.3$-$2528    &   1379  & 2.48 $\pm$ 0.05 $\pm$ 0.03 &  2.03 $\pm$ 0.11 $\pm$ 0.06              &  80.84  & $-$25.48   &  no/no                      & ...                  & ...             \\
PSR J1023+0038                           & ...              &   1094    & 2.37 $\pm$ 0.03 $\pm$ 0.08 &  1.96 $\pm$ 0.01 $\pm$ 0.13      & 155.92  &  0.68    &  no/no\tablenotemark{a}     &   1.90 $\pm$ 0.01   $\pm$ 0.13  &    0.90            \\
PSR J1023+0038$\dag$      & ...              &   76    & 2.31 $\pm$ 0.03 $\pm$ 0.04 &  0.50 $\pm$ 0.09 $\pm$ 0.05              & 155.92  &  0.68    &  no/no     &   0.48 $\pm$ 0.09   $\pm$ 0.05  &    0.90            \\
PSR J1023+0038$\ddag$     &...               &   1653  & 2.41 $\pm$ 0.10 $\pm$ 0.13 &  5.55 $\pm$ 0.20 $\pm$ 0.27              & 155.92  &  0.68    &  no/no                      &   5.38 $\pm$ 0.19   $\pm$ 0.25  &    0.90      \\

PSR J1227$-$4853   &   J1227.9$-$4854   &   2449   & 2.40 $\pm$ 0.02 $\pm$ 0.06 &  3.72 $\pm$ 0.11 $\pm$ 0.07        & 186.98  & $-$48.90   &  no/yes                     &   17.80 $\pm$ 0.53  $\pm$ 0.34  &  2.00           \\

PSR J1227$-$4853$\dag$   &   J1227.9$-$4854   &   2059   & 2.36 $\pm$ 0.06 $\pm$ 0.09 &  4.57 $\pm$ 0.15 $\pm$ 0.08   & 186.98  & $-$48.90   &  no/yes                     &   21.85 $\pm$ 0.70  $\pm$ 0.42  &  2.00           \\
PSR J1227$-$4853$\ddag$  &   J1227.9$-$4854   &   370    & 2.42 $\pm$ 0.03 $\pm$ 0.15 &  1.79 $\pm$ 0.16 $\pm$ 0.17                   & 186.98  & $-$48.90   &  no/yes                     & 8.57 $\pm$ 0.75     $\pm$ 0.82  &   2.00                   \\
PSR J1431$-$4715	      &...    	         &   13    & ... &  $        < 0.45             $                   & 217.75  &  $-$47.25  &  no/yes                      &    $<$ 1.36          &    2.42       \\
3FGL J1544.6$-$1125        &   J1544.6$-$1125   &   407   & 2.48 $\pm$ 0.05 $\pm$ 0.11 &  1.43 $\pm$ 0.09 $\pm$ 0.04                 & 236.17   & $-$11.43   &  no/no                      & ...                  & ...                 \\
PSR J1628$-$3205 	      &    J1628.0$-$3203  &   372   & 2.36 $\pm$ 0.02 $\pm$ 0.04 &  1.01 $\pm$ 0.09 $\pm$ 0.09                      & 247.02  & $-$32.06   &  no/yes                      &   2.87 $\pm$ 0.25   $\pm$  0.26 &    1.54        \\
3FGL J1653.6$-$0158	      &    J1653.6$-$0158  &   2911  & 2.24 $\pm$ 0.02 $\pm$ 0.07 &  3.28 $\pm$ 0.12 $\pm$ 0.05                        & 253.42  &  $-$1.98   &  no/no                      & ...                  & ...                \\
PSR J1723$-$2837	          &...    	         &   35    & 2.67 $\pm$ 0.15 $\pm$ 0.17 &  0.83 $\pm$ 0.23 $\pm$ 0.40                         & 260.75  & $-$28.62   &  no/no                      &    0.56  $\pm$ 0.16  $\pm$ 0.15 &    0.75             \\
PSR J1816+4510	          &    J1816.5+4512  &   952   & 2.12 $\pm$ 0.04 $\pm$ 0.02 & 0.94 $\pm$ 0.06 $\pm$ 0.04                   & 274.13  & 45.20    &  no/yes                     &   19.92 $\pm$ 1.22   $\pm$ 0.81 &    4.20       \\
PSR J2129$-$0429          &    J2129.6$-$0427  &   402   & 2.22 $\pm$ 0.05 $\pm$ 0.06 &  1.10 $\pm$ 0.08 $\pm$ 0.03                   & 322.41  &$-$4.46     &  no/yes                      &   1.40 $\pm$ 0.10    $\pm$ 0.04 &    1.03        \\
PSR J2215+5135	          &    J2215.6+5134  &   788   & 2.08 $\pm$ 0.04 $\pm$ 0.09 &  1.33 $\pm$ 0.09 $\pm$ 0.06               & 333.91  & 51.58    &  yes/yes                    &   17.32 $\pm$ 1.12   $\pm$ 0.75 &    3.30    \\
PSR J2339$-$0533 	      &    J2339.6$-$0533  &   3963  & 1.94 $\pm$ 0.01 $\pm$ 0.04 &  5.05 $\pm$ 0.19 $\pm$ 0.53               & 354.90  & $-$5.55    &  no/yes                     &   7.30 $\pm$ 0.27    $\pm$ 0.77 &    1.10      \\
\hline
BW name               &                  &            &                   &                               &                          &          &                                      &                      &                    \\
\hline
PSR B1957+20              &    J1959.5+2047  &   562   & 2.40 $\pm$ 0.04 $\pm$ 0.38 & 1.52 $\pm$ 0.02 $\pm$ 0.06           &  299.89 &  20.80   &  yes/yes                    &   4.05 $\pm$ 0.33  $\pm$ 0.38   &    1.53         \\
PSR J0023+0923            &   J0023.4+0923   &   407  & 2.27 $\pm$ 0.01 $\pm$ 0.08 &  0.79 $\pm$ 0.07 $\pm$ 0.13                  &  5.86   &  9.39    &  yes/yes                    &   0.85 $\pm$ 0.07  $\pm$ 0.14   &    0.95      \\
PSR J0610$-$2100          &    J0610.2$-$2059  &   366   & 2.28 $\pm$ 0.05 $\pm$ 0.04 &  1.09 $\pm$ 0.08 $\pm$ 0.04                  &  92.55  & $-$20.99   &  yes/yes                    &   41.44 $\pm$ 2.97 $\pm$ 1.65   &    5.64       \\
PSR J1124$-$3653          &    J1123.9$-$3653  &   1041  & 2.12 $\pm$ 0.04 $\pm$ 0.07 & 1.28 $\pm$ 0.08 $\pm$ 0.03               &  170.99 &  $-$36.89  &  yes/yes                    &   29.60 $\pm$ 1.88 $\pm$ 0.57   &    4.40    \\
PSR J1301+0833              &    J1301.6+0832  &   505   & 2.25 $\pm$ 0.05 $\pm$ 0.08 &  1.18 $\pm$ 0.08 $\pm$ 0.13                &  195.42 &  8.54    &  no/yes                     &   1.17 $\pm$ 0.08  $\pm$ 0.13   &    0.91       \\
PSR J1311$-$3430          &    J1311.8$-$3430  &   9523  & 2.21 $\pm$ 0.01 $\pm$ 0.04 &  6.20 $\pm$ 0.13 $\pm$ 0.04               &  197.96 &  $-$34.50  &  no/yes                     &   102.72 $\pm$ 2.13  $\pm$ 0.78 &    3.72       \\
PSR J1446$-$4701          &    J1446.6$-$4701  &   163   & 2.08 $\pm$ 0.03 $\pm$ 0.05 &  0.91 $\pm$ 0.08 $\pm$ 0.07               &  221.66 &  $-$47.03  &  yes/yes                    &   4.47 $\pm$ 0.41    $\pm$ 0.35 &    2.03       \\
PSR J1544+4937            &    J1544.0+4938  &   156   & 2.27 $\pm$ 0.09 $\pm$ 0.09 &  0.45 $\pm$ 0.05 $\pm$ 0.02                  &  236.02 &  49.65   &  no/yes                     &   2.85 $\pm$ 0.31    $\pm$ 0.14 &    2.30       \\
PSR J1653$-$2054          &...               &   1     & 1.47 $\pm$ 0.73 $\pm$ 0.16 &  $     < 0.14               $        &  253.38 &  $-$20.92  &  no/no                     &      $<$ 0.53           &    2.64      \\
PSR J1731$-$1847          &...               &   6      & 2.32 $\pm$ 0.20 $\pm$ 0.38 &  $     < 0.55               $              &  262.82 &  $-$18.79  &  no/no                      &      $<$ 4.65           &    4.03         \\
PSR J1745+1017            & ...              &   0.00     & 2.38 $\pm$12.64 $\pm$ 0.32 &  $     < 0.17              $              &  266.25 &  10.28   &  no/yes                      &      $<$ 0.12           &    1.36        \\
PSR J1810+1744            &    J1810.5+1743  &   1701  & 2.35 $\pm$ 0.03 $\pm$ 0.43 &  2.34 $\pm$ 0.09 $\pm$ 0.18               &  272.64 &  17.72   &  yes/yes                    &   17.38 $\pm$ 0.69   $\pm$ 1.37 &    2.49     \\
PSR J2047+1053            &    J2047.1+1054  &   140   & 2.34 $\pm$ 0.53 $\pm$ 0.12 &  0.29 $\pm$ 0.03 $\pm$ 0.12              &  311.78 &  10.91   &  yes/yes                    &   1.74 $\pm$ 0.20    $\pm$ 0.68 &    2.23     \\
PSR J2051$-$0827          &    J2051.3$-$0828  &   126   & 2.21 $\pm$ 0.10 $\pm$ 1.09 &  0.33 $\pm$ 0.05 $\pm$ 0.15           &  312.83 &  $-$8.48   &  yes/yes                    &   0.64 $\pm$ 0.11    $\pm$ 0.16 &    1.28       \\
PSR J2214+3000            &    J2214.6+3000  &   5380  & 2.04 $\pm$ 0.02 $\pm$ 0.05 &  3.16 $\pm$ 0.09 $\pm$ 0.01                &  333.66 &  30.01   &  yes/yes                    &   6.58 $\pm$ 0.18    $\pm$ 0.03 &    1.32      \\
PSR J2234+0944            &    J2234.8+0945  &   720   & 2.20 $\pm$ 0.05 $\pm$ 0.10 &  1.14 $\pm$ 0.09 $\pm$ 0.10                  &  338.71 &   9.75   &  no/yes                     &   1.45 $\pm$ 0.11    $\pm$ 0.13 &    1.03        \\
PSR J2241$-$5236          &    J2241.6-5237  &   6424  & 2.03 $\pm$ 0.02 $\pm$ 0.07 &  3.27 $\pm$ 0.09 $\pm$ 0.03                 &  340.42 & $-$52.62   &  yes/yes                    &   1.81 $\pm$ 0.05    $\pm$ 0.02 &    0.68    \\
PSR J2256$-$1024          &    J2256.7$-$1022  &   439   & 2.02 $\pm$ 0.05 $\pm$ 0.05 &  0.56 $\pm$ 0.05 $\pm$ 0.11              &  344.18 &  $-$10.38  &  no/no                      &   0.55 $\pm$ 0.05    $\pm$ 0.11 &    0.91       \\
\hline
\end{tabular}
\tablenotetext{a}{Gamma-ray pulsations from J1023+0038
	were detected
	by Archibald et al. (2013); pulsations from
	J1227$-$4853 were detected by Johnson et al. (2015).
}
\end{table}
\end{landscape}

\begin{landscape}
	\begin{table}[hbt!]
		\scriptsize
		\centering
		\caption{Long-term {light curve} {fit results}. Columns indicate
		the binning, the $
		\chi^2$ of a constant fit (except for J1023+0038 in the whole time period and before the transition, see Fig. \ref{odd}, and other cases
		where only upper limits are obtained)
		and the number of degrees of freedom, its $P-$value and $\sigma$. The last two columns {(jump factor \& data points)} represent
		the features of the transitions that our analysis can rule out {(see section 3.2 for {details})}. {$\dag$ and $\ddag$ indicate the data period
        before and after the state transition. Ellipses are used when a value is not available.} In the case of J2234+0944
		we also present the fittings before ($\dag$) and after ($\ddag$) the jump visible in Fig. \ref{odd}. For details, see text.}
		\vspace{0.1cm}
		\label{table2}
		\begin{tabular}{llllccccc}
\hline
RB name                                        &   Binning &$\chi^2$ & dof & $P-value$ & $\sigma$  & Jump & Data\\
     PSR                                            & (days)                  &                 &       &                  &                 & factor & points \\		
		\hline
3FGL J0523.3$-$2528                    & 66.9    &  52.6  &   36  &  $3.7 \times 10^{-2}$    &   2.1	&  1.74   &   2      \\
PSR J1023+0038                             &	9.3      & ...&...	&...	    &...  	&...	  &...	    \\
PSR J1023+0038$\dag$             &	9.3      & ... &... 	&... 	     &...  	&...	  &...	    \\
PSR J1023+0038$\ddag$     &	 9.3     & 67.0 	& 62	&	0.3     & 1.0 	&	...  &...	\\
PSR J1227$-$4853        &	53.5   & 205.7	&	44 &  $\ll 10^{-10}$   & 9.8 	&	...  &...	      \\
PSR J1227$-$4853$\dag$    &53.5	 & 77.1  	& 29	 &	 $3.1 \times 10^{-6}$     & 4.7 	&	...  &	...      \\
PSR J1227$-$4853$\ddag$     &53.5	  & 13.0	& 14	&	0.5     & 0.6 	&	1.87  &	4	    \\
PSR J1431$-$4715	        	          &  ... 	&	...  &	 ...  &	  ...    & ... 	 &	...     & ...  \\
3FGL J1544.6$-$1125        & 214.8   &  6.6	&   10  &   0.8 &   0.3	&  1.63   &    4\\
PSR J1628$-$3205 	       & 833.8   &  3.9	&   2	   &   0.1 &   1.5	&  1.47   &    1   \\
3FGL J1653$-$0158	       &  36.5   & 73.3	&   67   &   0.3   &   1.1	&  1.74   &   5 	\\
PSR J1723$-$2837	   	            &  368.0  & 1.1   &   1   &   0.3   &   1.0   &   2.18   &   2  	    \\
PSR J1816+4510	     &   94.0 & 23.2   &   24	&   0.5   &	0.7	&   1.77   &   4   \\
PSR J2129$-$0429     &   311.7 & 5.4   &   7   &   0.6   &   0.5   &  1.60   &   4	    \\
PSR J2215+5135	  &   138.6 & 10.1	&   17   &   0.9   &   0.1	& 1.68   &   5      \\
PSR J2339$-$0533 	      & 24.5     & 128.0 &  96	 & $1.6 \times 10^{-2}$  &  2.4	&  1.91   &    2\\
\hline
BW name               		\\
\hline
PSR B1957+20            &   192.0 & 9.8   &   12   &   0.6   &   0.5   &    1.60  &	4	 \\
PSR J0023+0923            &   156.0 & 7.5   &   14   &   0.9   &   0.1   &    1.81  &	5	\\
PSR J0610$-$2100      &   171.1 & 7.0   &   14   &   0.9   &   0.1   &    1.80  &	5	    \\
PSR J1124$-$3653             &   104.4 & 19.3  &   23   &   0.7   &   0.4   &    1.72  &	5	\\
PSR J1301+0833              &   126.4 & 31.7  &   17   &   $1.7 \times 10^{-2}$   &   2.4	 &   1.85   &	1	   \\
PSR J1311$-$3430        &   11.8  & 183.3 &   203  &   0.8   &   0.2   &   1.76   &	11	    \\
PSR J1446$-$4701         &   506.4 & 26.3  &   4    &   $2.7 \times 10^{-5}$ &   4.2	 &   2.09   &...		 \\
PSR J1544+4937              &   370.2 & 2.5   &   6    &   0.9   &   0.2   &   1.85   &	5	    \\
PSR J1653$-$2054        &  ... 	&	...  &	 ...  &	  ...    & ... 	 &	...     & ...  \\
PSR J1731$-$1847          &  ... 	&	...  &	 ...  &	  ...    & ... 	 &	...     & ...  \\
PSR J1745+1017             &  ... 	&	...  &	 ...  &	  ...    & ... 	 &	...     & ...  \\
PSR J1810+1744            &  48.7	& 63.9  &   48   &   $6.2 \times 10^{-2}$   &   1.9   &   1.75   &	3      \\
PSR J2047+1053           & 445.6	& 4.4   &   5    &   0.5   &   0.7   &   1.99   &	3	  \\
PSR J2051$-$0827         & 688.6	&  1.4  &   3    &   0.7   &   0.4   &   1.79   &	4	     \\
PSR J2214+3000           &  17.9	& 116.4 &   131  &   0.8   &   0.2   &   1.85   &	9	   \\
PSR J2234+0944               &  106.7  & 77.8  &   22   & $3.7 \times 10^{-8}$  &   5.5   &   1.82   &	...       \\
PSR J2234+0944$\dag$     &106.7    	& 9.5   &   7    &  0.2 &  1.2	 &   1.84   &	2\\
PSR J2234+0944$\ddag$    &106.7    	& 17.4  &   14   &  0.2 &  1.2	 &   1.66   &	 3	      \\
PSR J2241$-$5236          &   15.2  & 138.8 &  153   & 0.8  &  0.3	 &   1.90   &	 10 \\
PSR J2256$-$1024          & 182.6	& 11.5  &  10    & 0.3  &  1.0	 &   1.77   &	 3    \\
\hline
\end{tabular}
\end{table}
\end{landscape}

\begin{figure}
\centering
\includegraphics[width=0.32\textwidth]{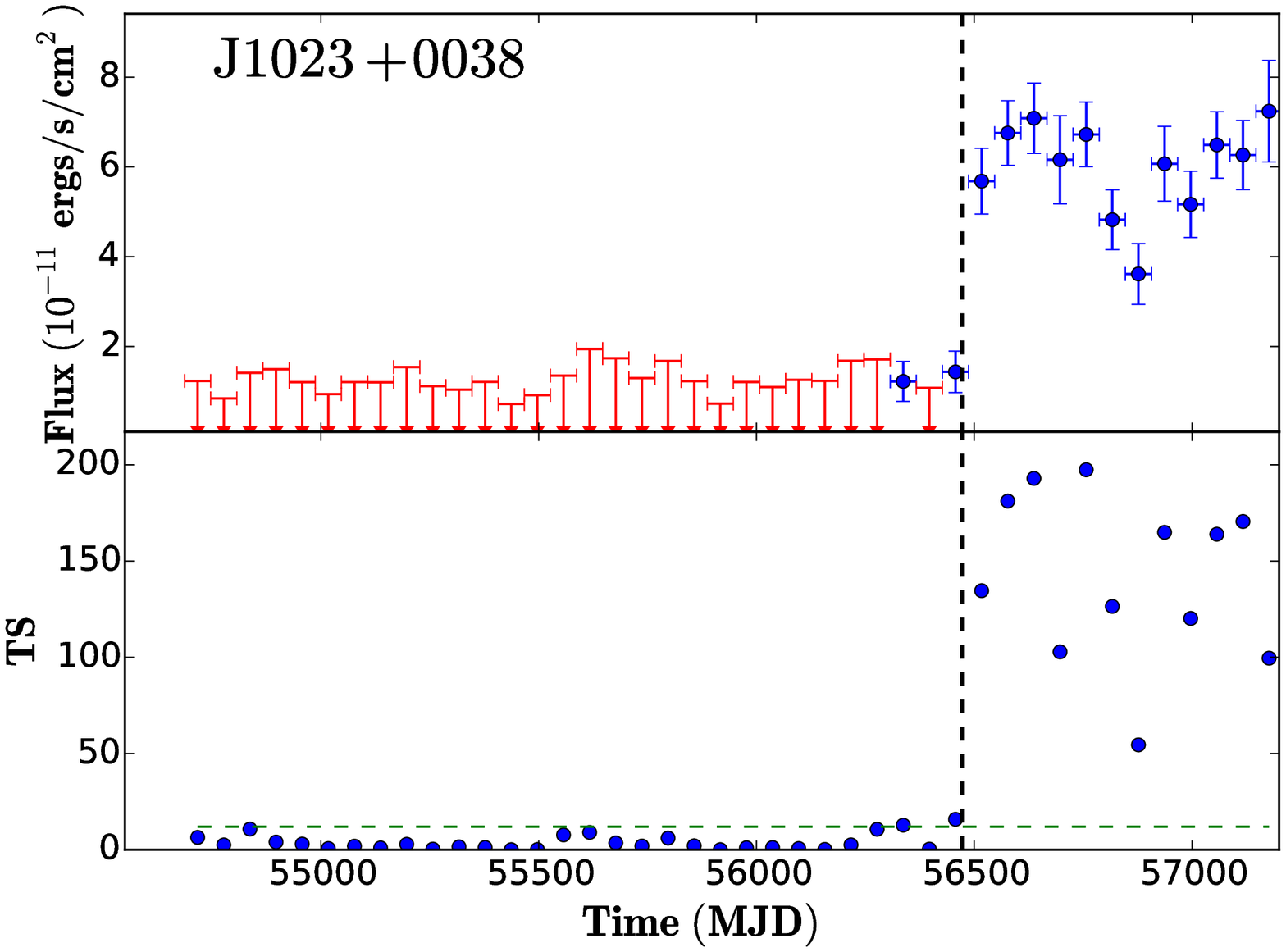}
\includegraphics[width=0.32\textwidth]{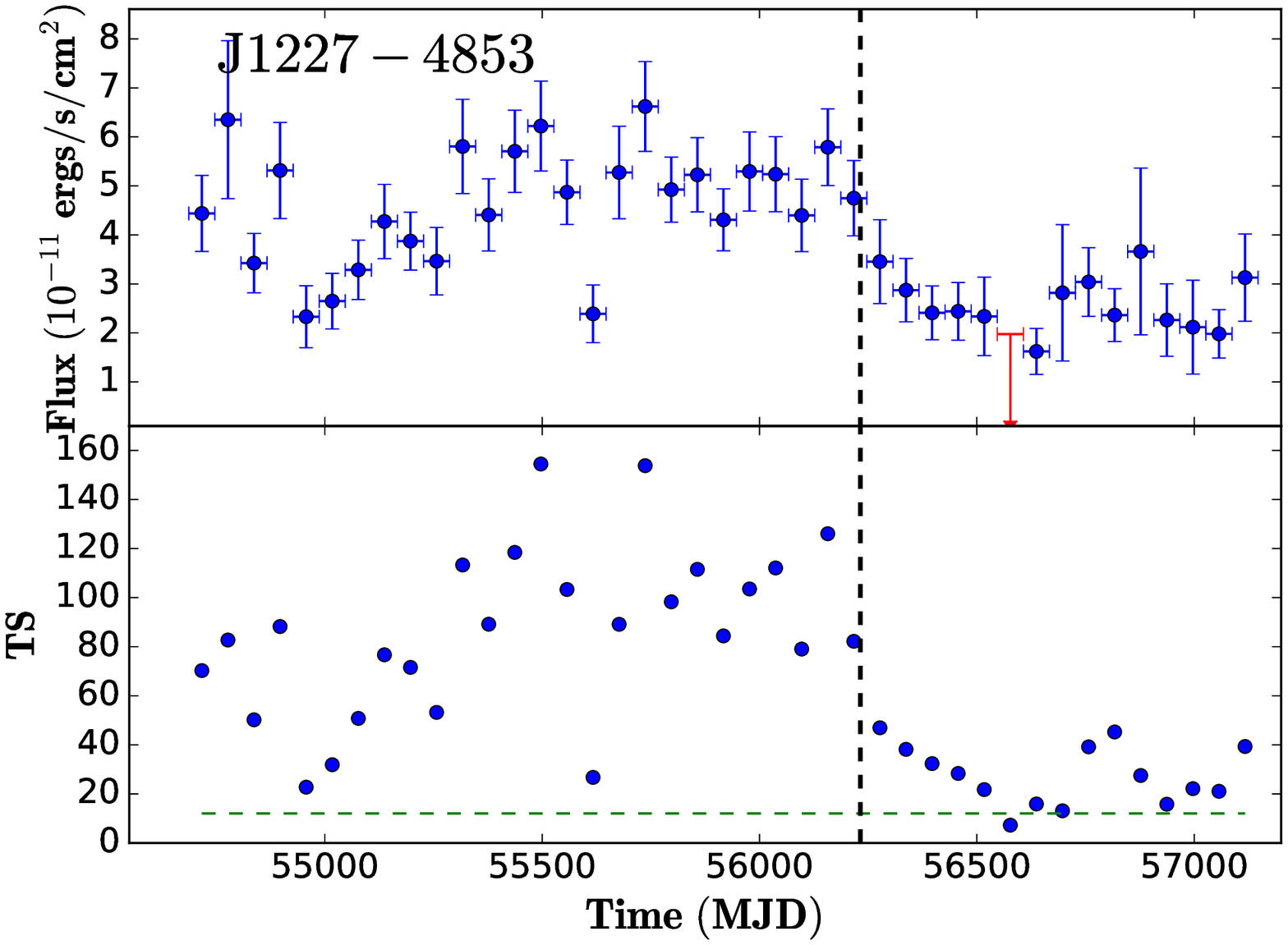}
\includegraphics[width=0.32\textwidth]{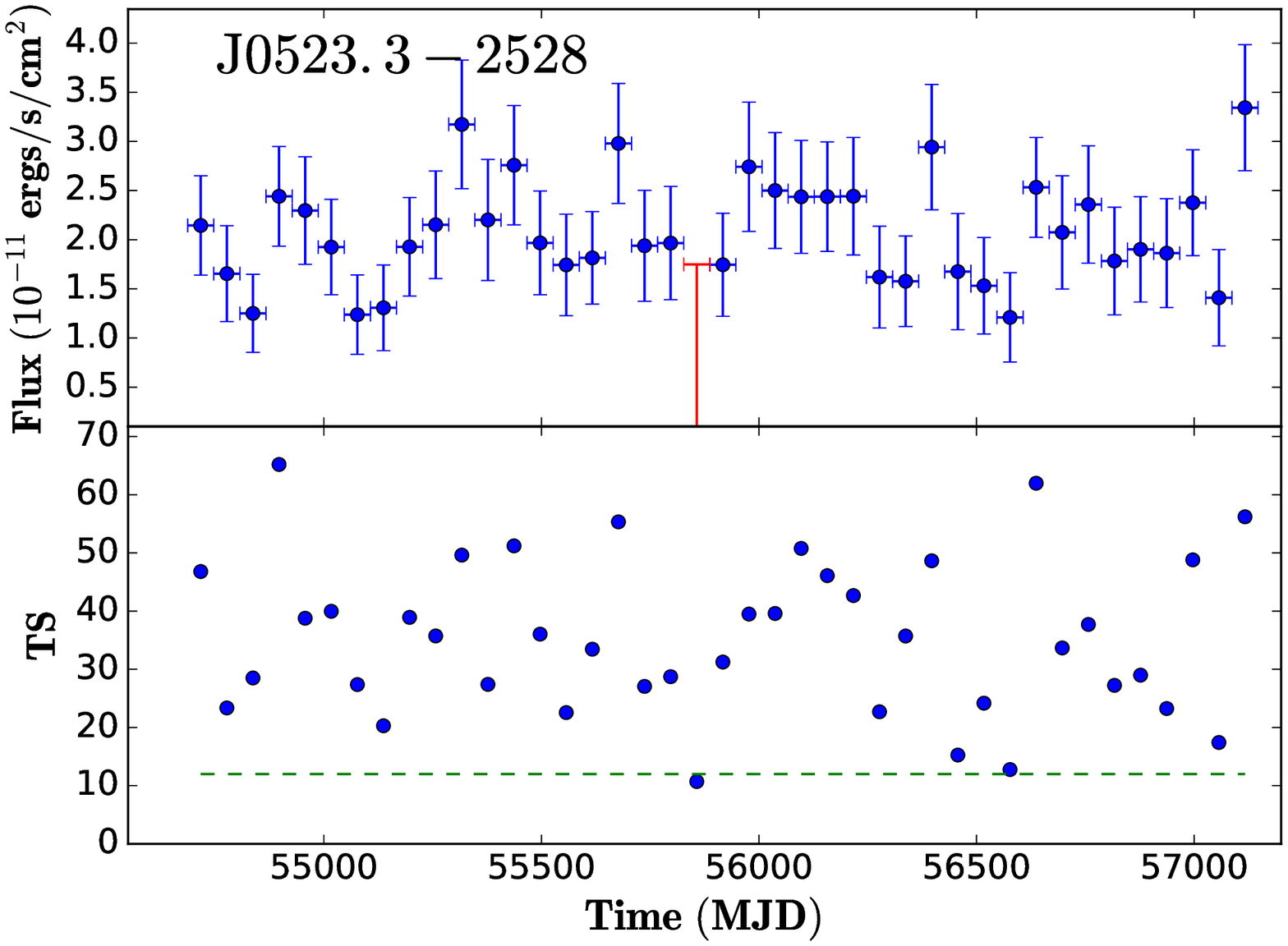}
\includegraphics[width=0.32\textwidth]{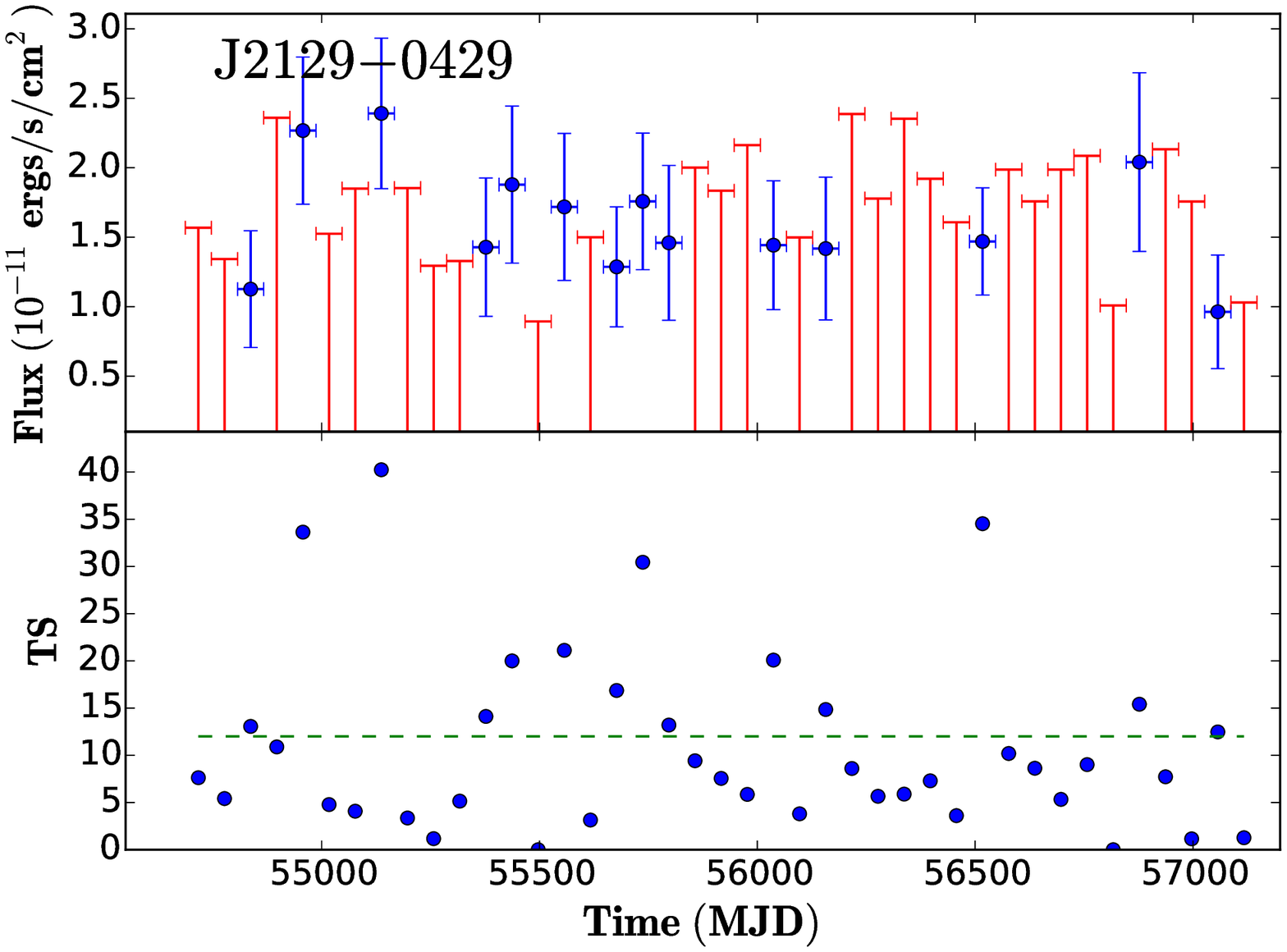}
\includegraphics[width=0.32\textwidth]{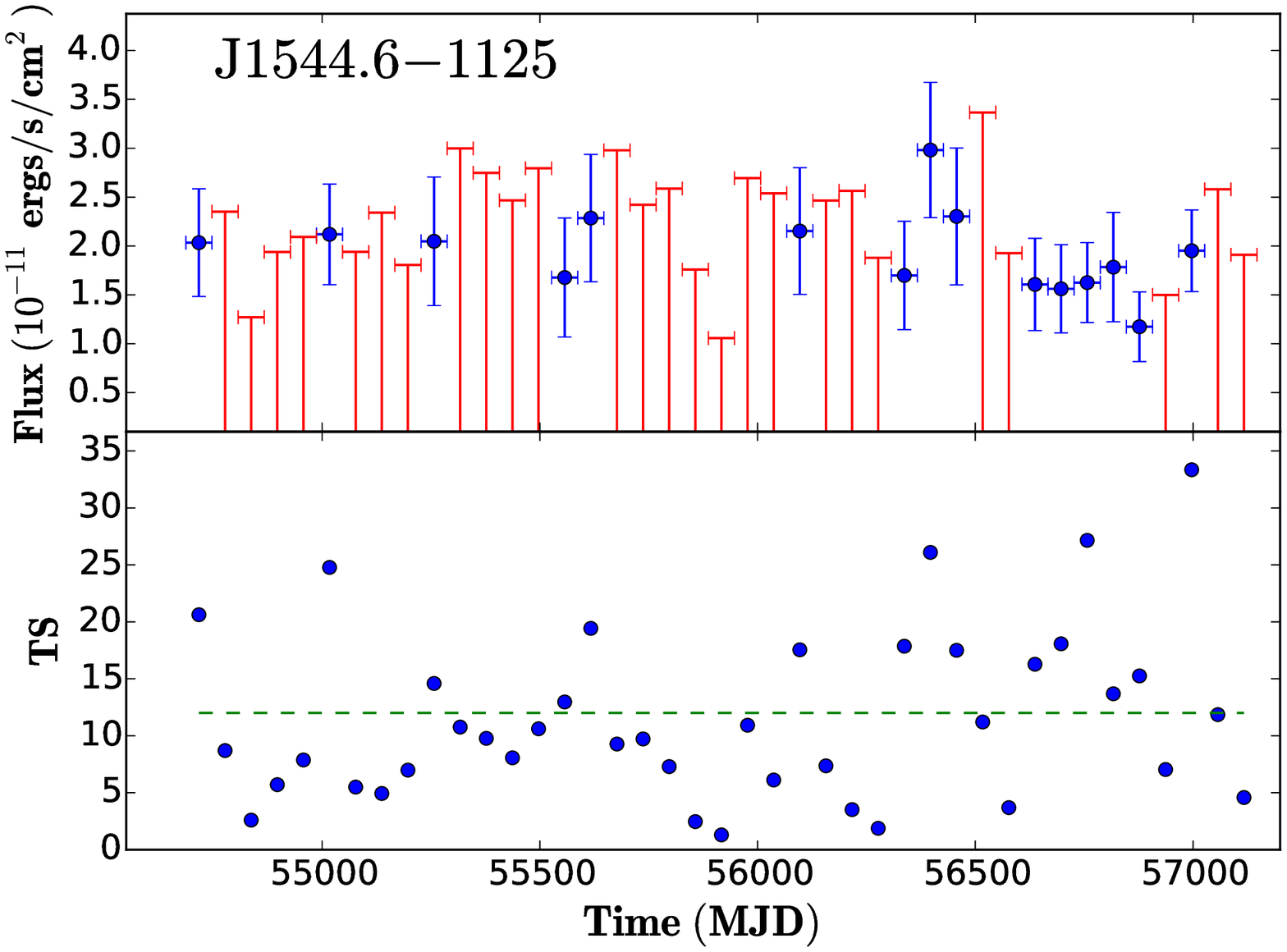}
\includegraphics[width=0.32\textwidth]{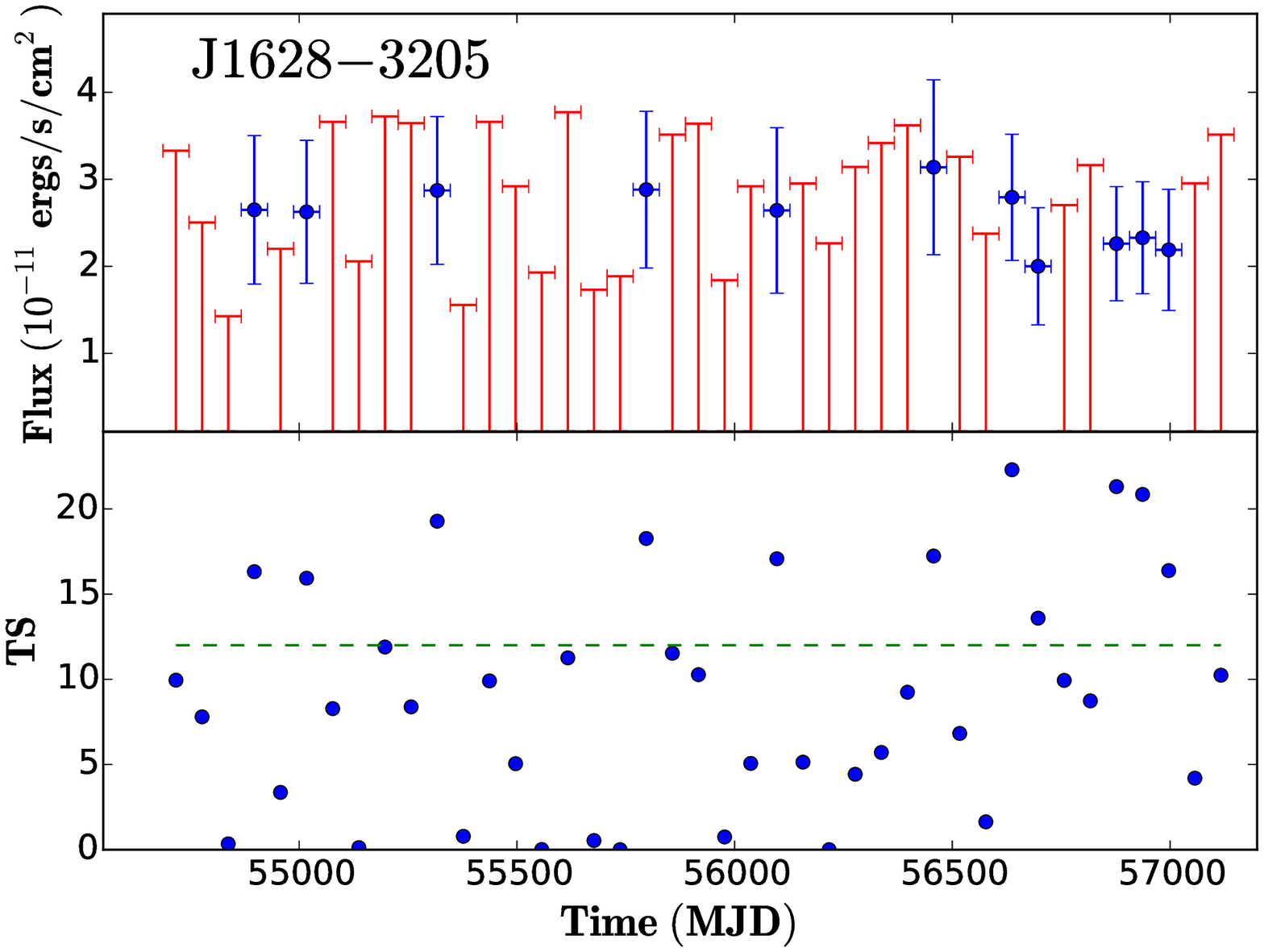}
\includegraphics[width=0.32\textwidth]{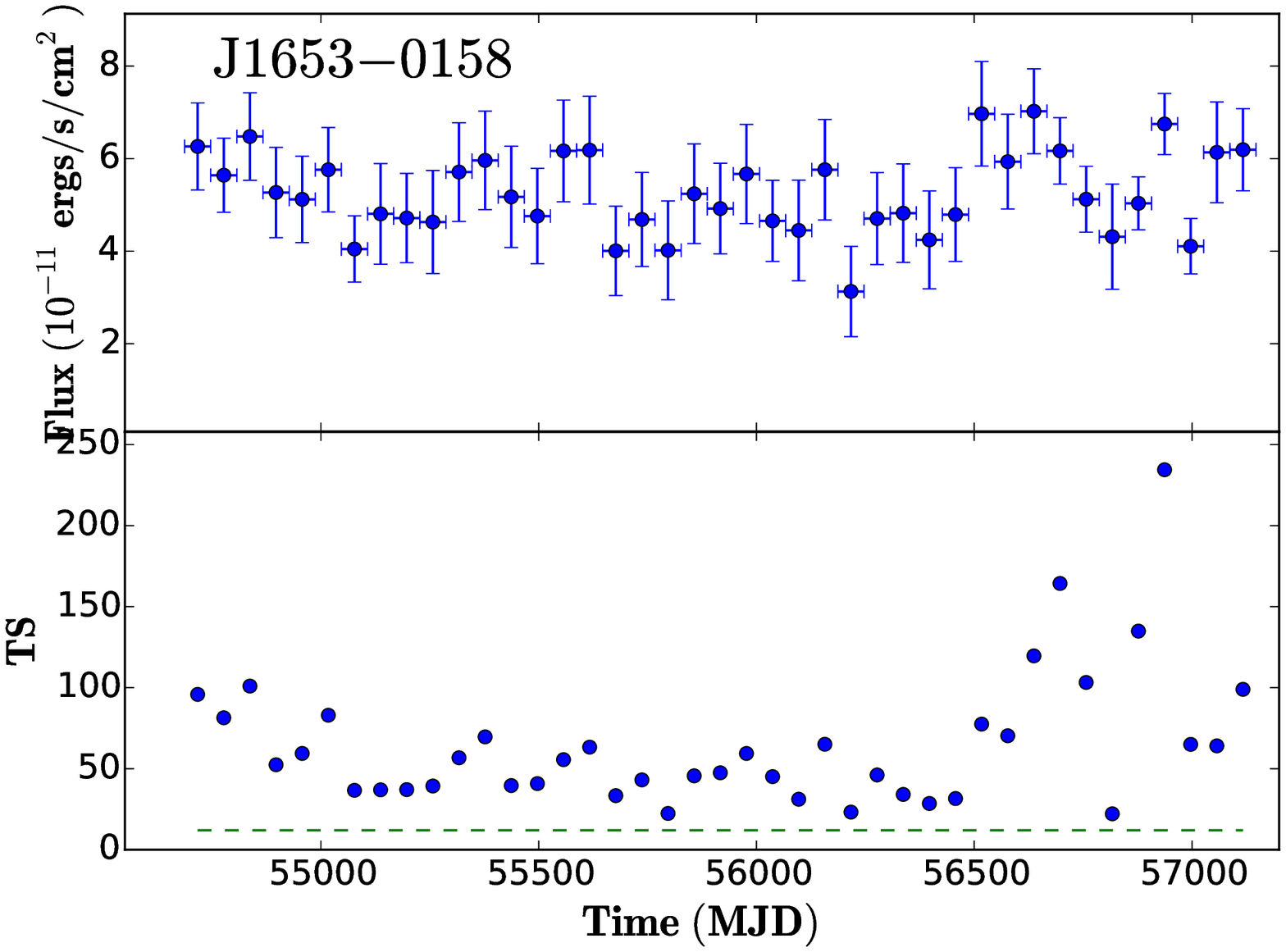}
\includegraphics[width=0.32\textwidth]{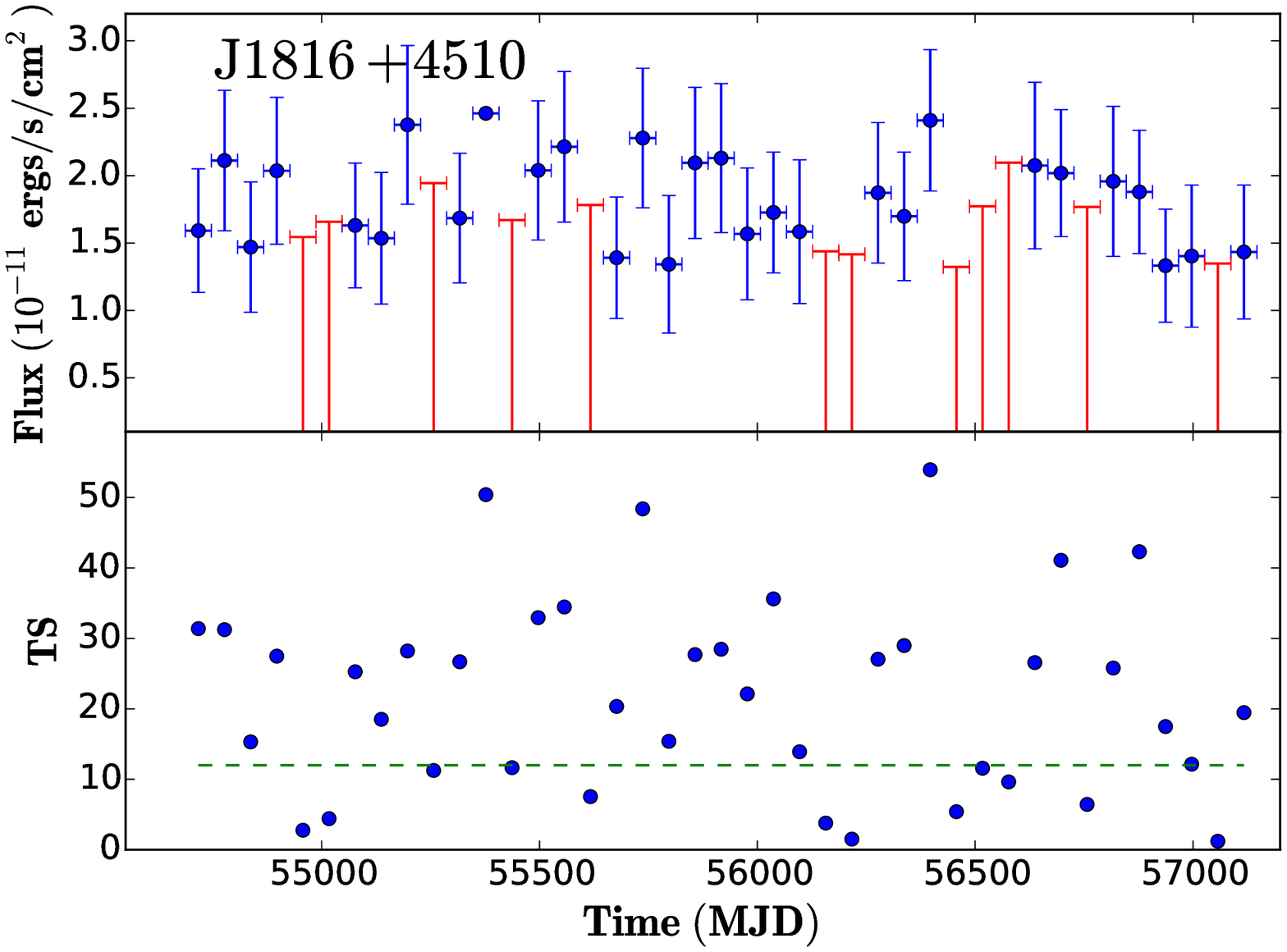}
\includegraphics[width=0.32\textwidth]{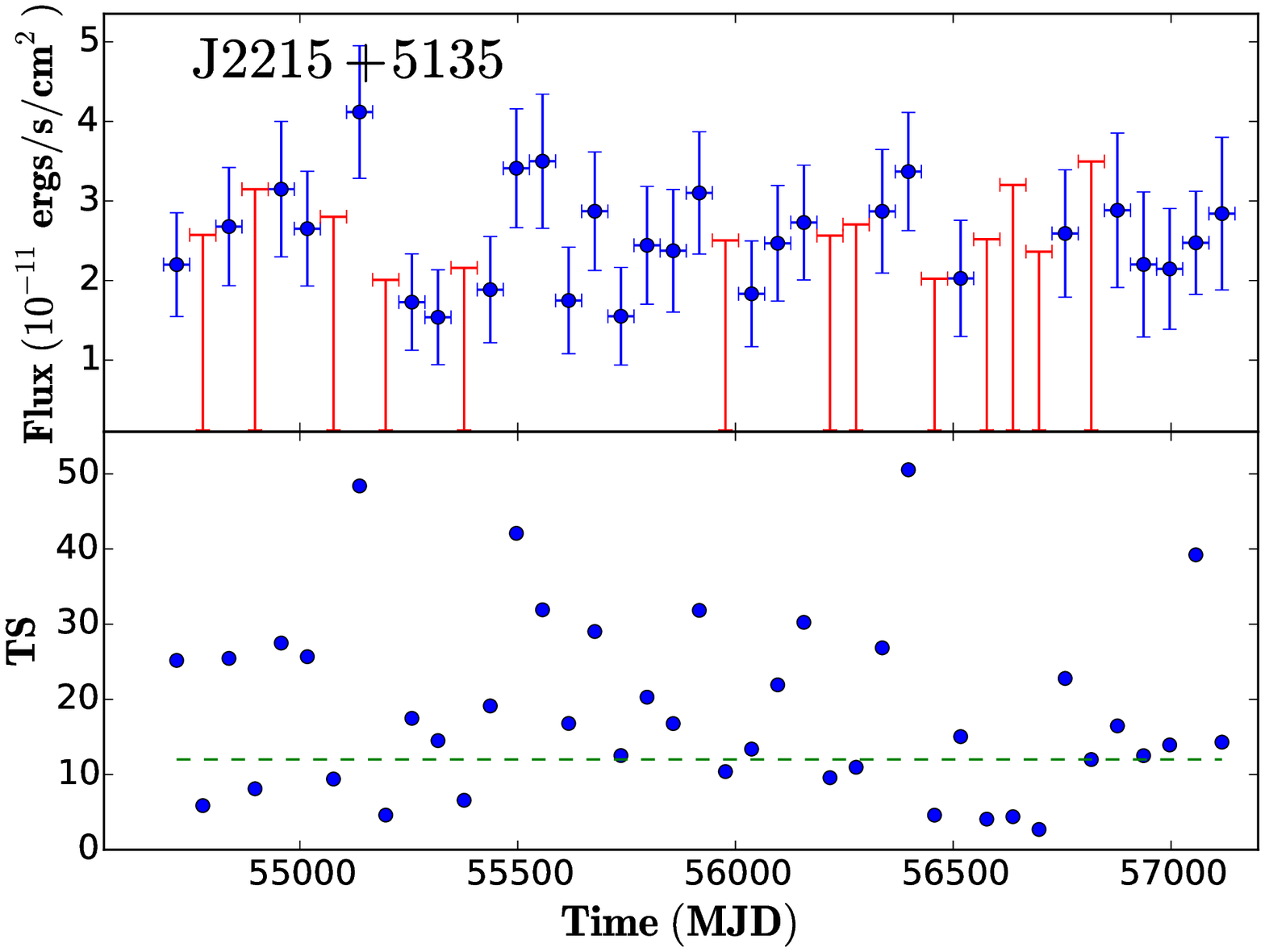}
\includegraphics[width=0.32\textwidth]{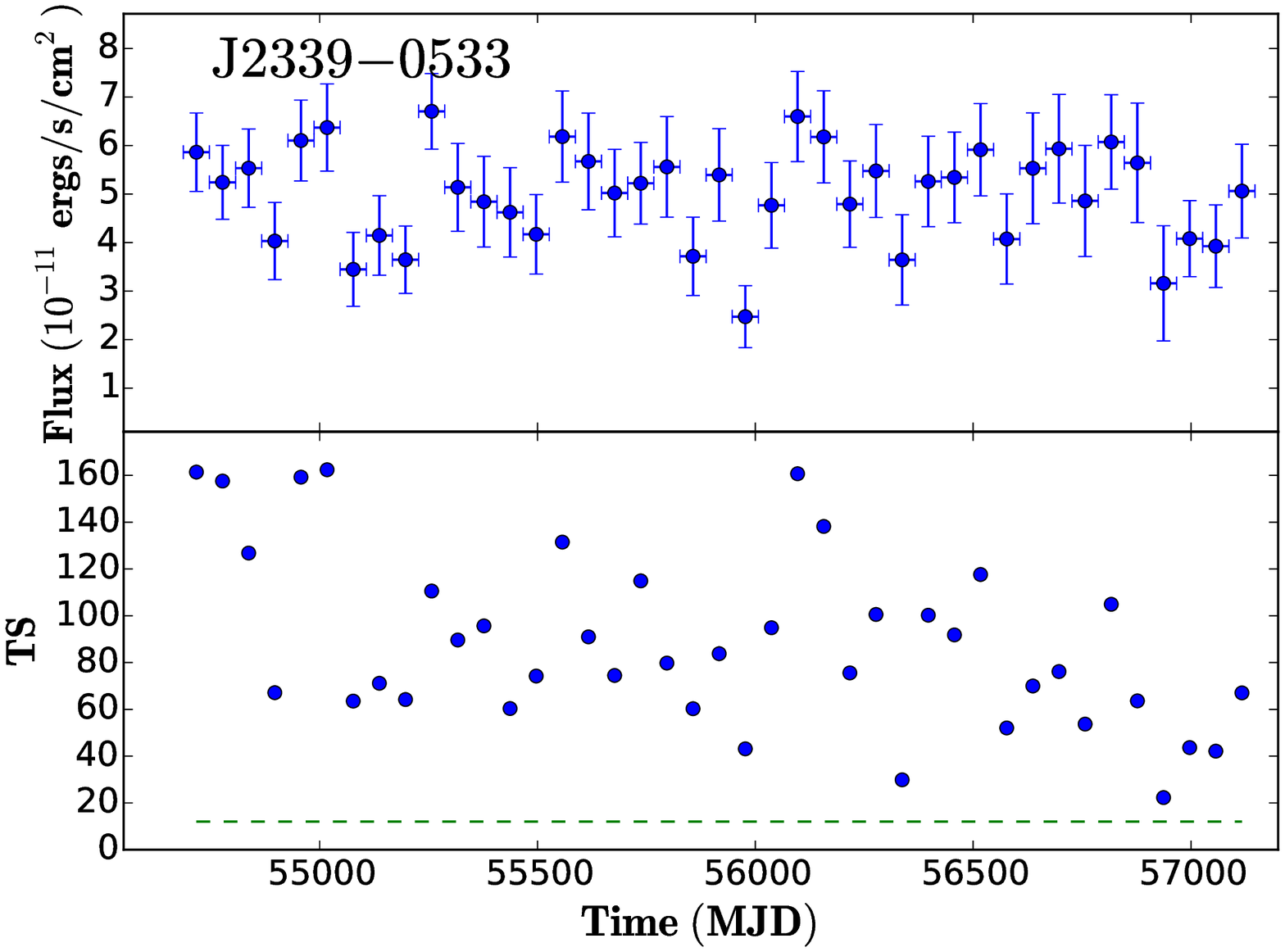}
\includegraphics[width=0.32\textwidth]{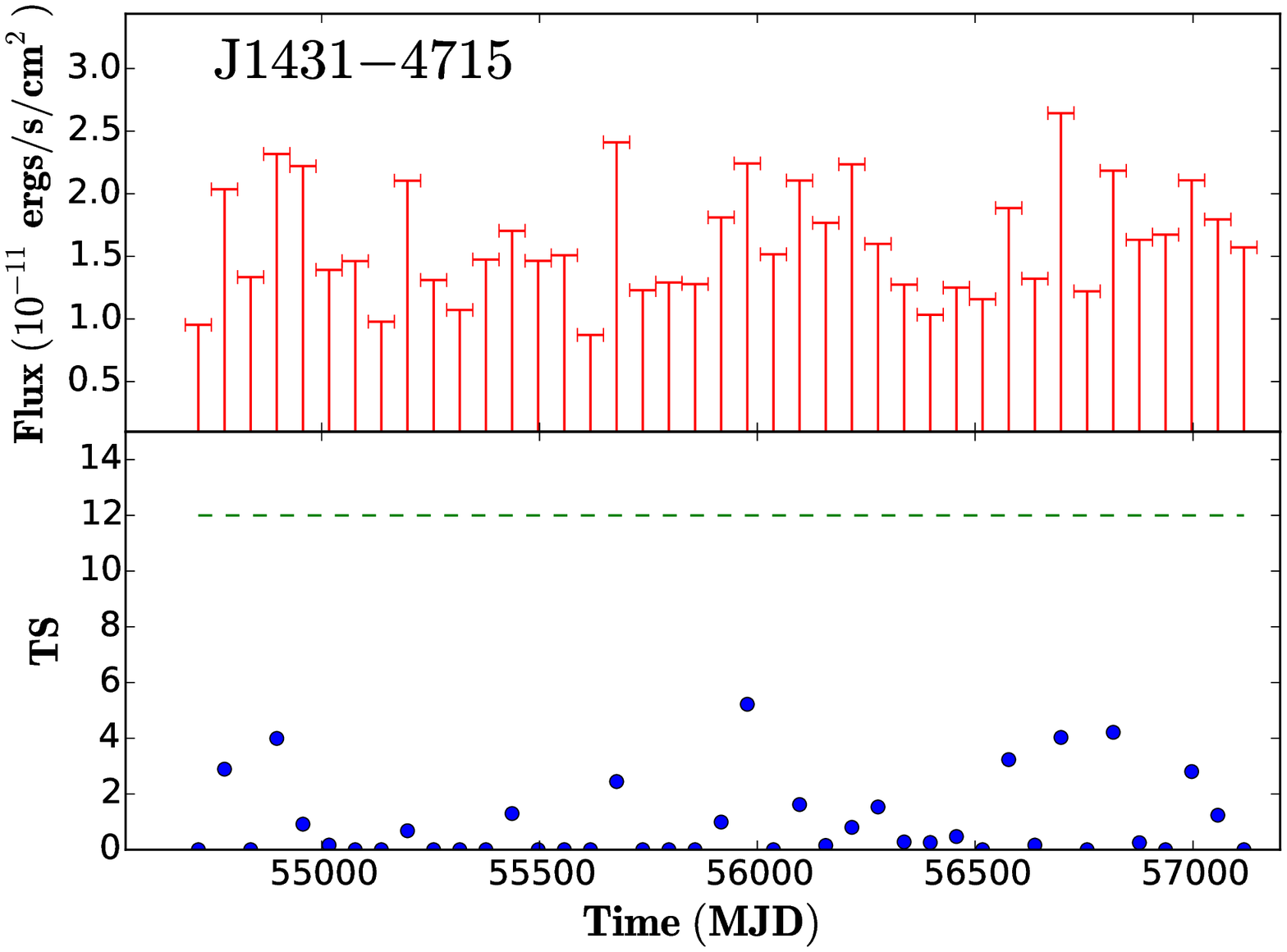}
\includegraphics[width=0.32\textwidth]{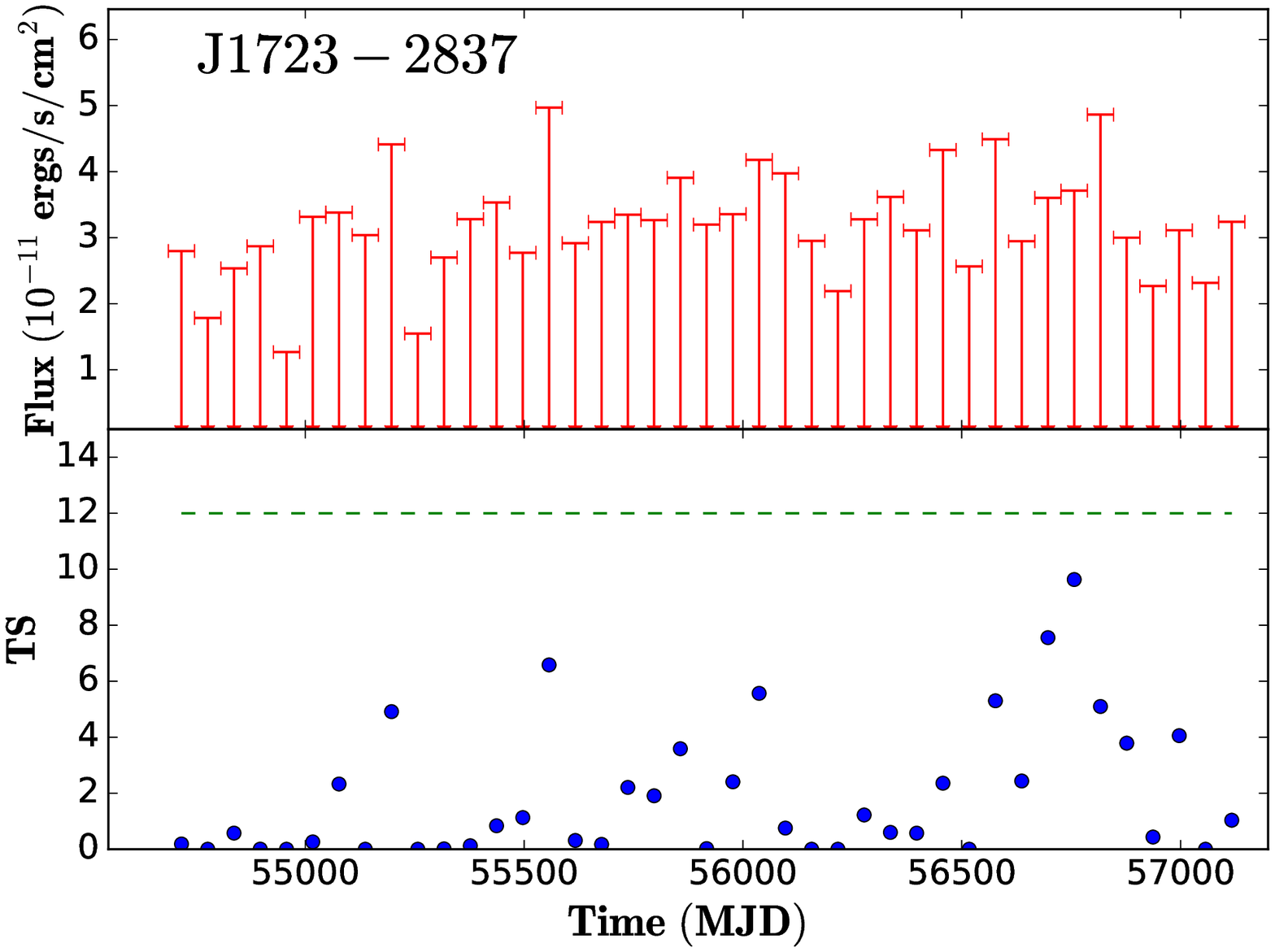}
\caption{First two panels: Long-term light curves of the transitional millisecond pulsars J1023+0038 and J1227$-$4853. The already known (Stappers et al. 2014; Bassa et al. 2014) state transitions are indicated with dotted vertical lines. The red lines show the flux upper limits. The dotted horizontal green line indicates TS=12. Subsequent panels: Long-term light curves of RBs and BWs studied (other than J1023+0038 and J1227$-$4853). The color coding remains the same. The time {binning} is 60 days in all cases.}
\label{LC}
\end{figure}

\addtocounter{figure}{-1} \begin{figure}[!tbh]
\centering
\includegraphics[width=0.32\textwidth]{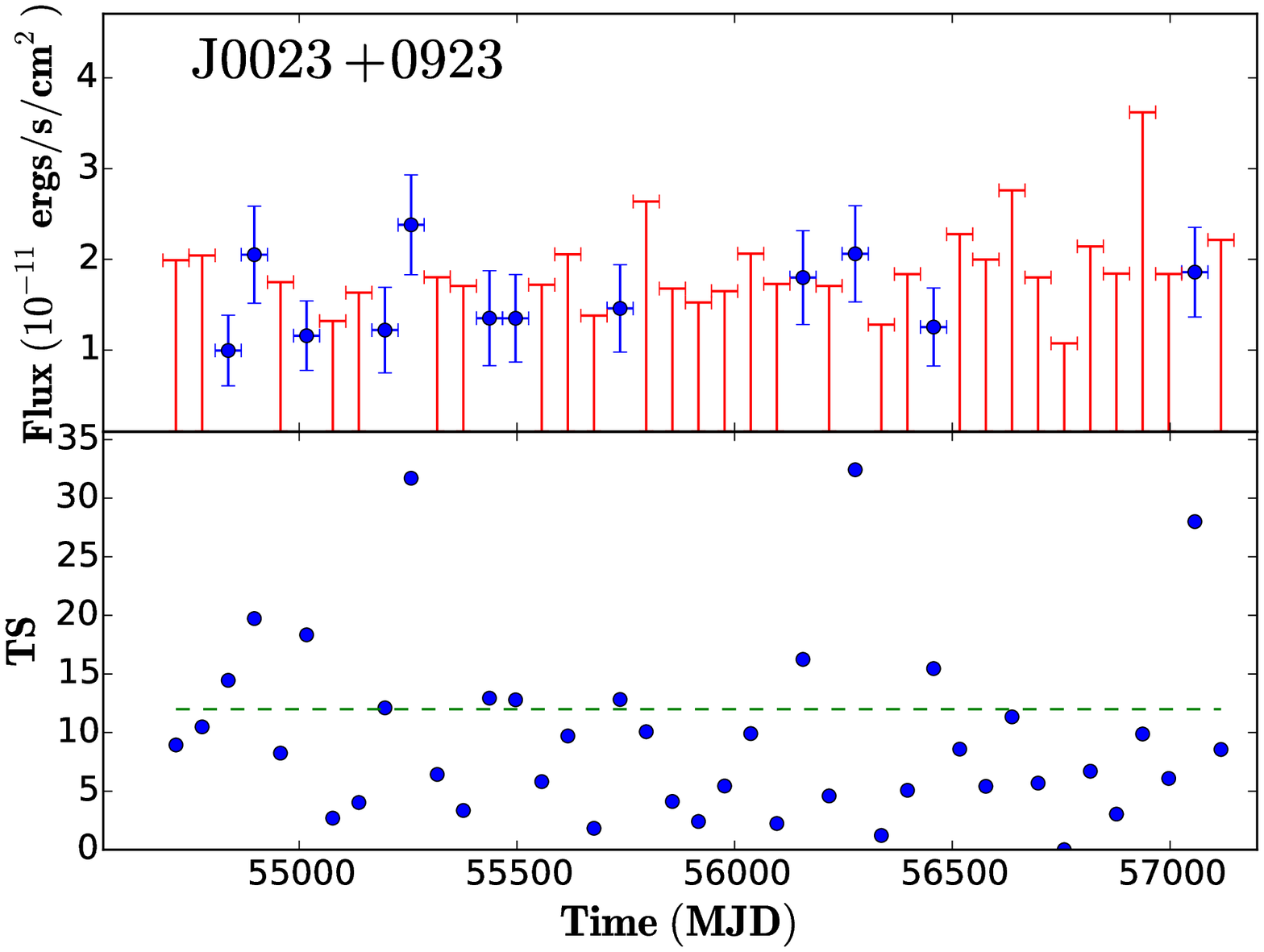}
\includegraphics[width=0.32\textwidth]{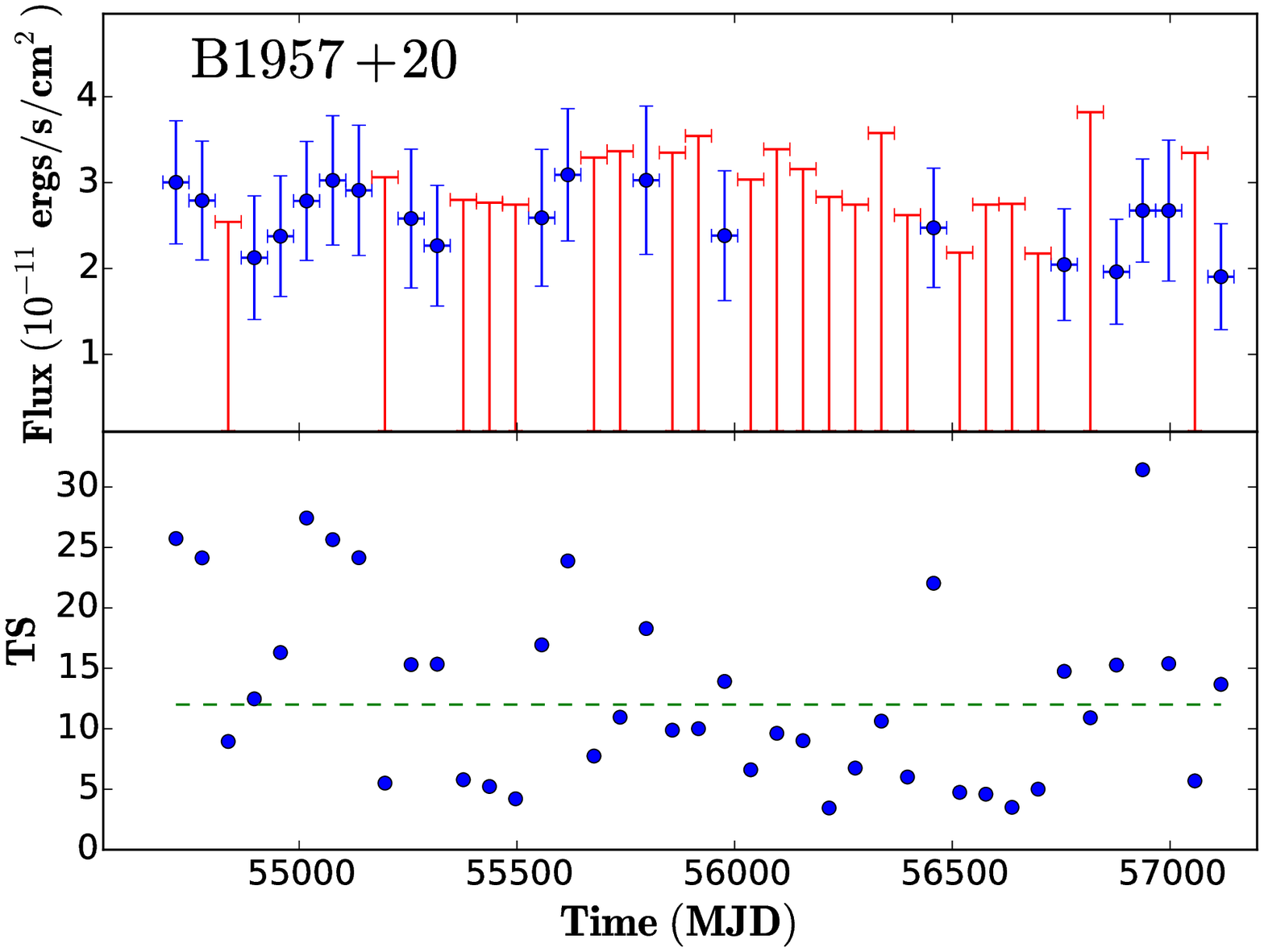}
\includegraphics[width=0.32\textwidth]{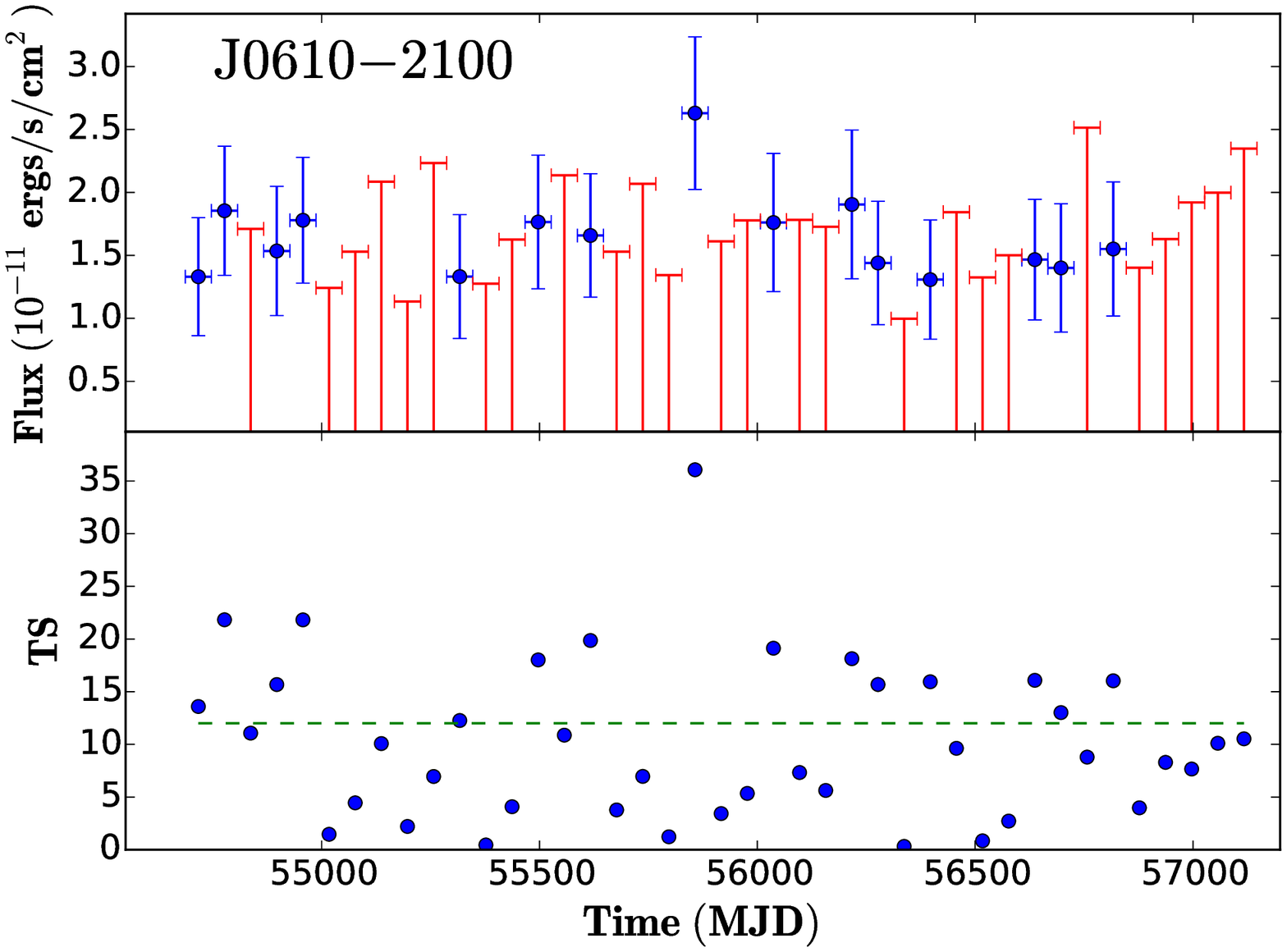}
\includegraphics[width=0.32\textwidth]{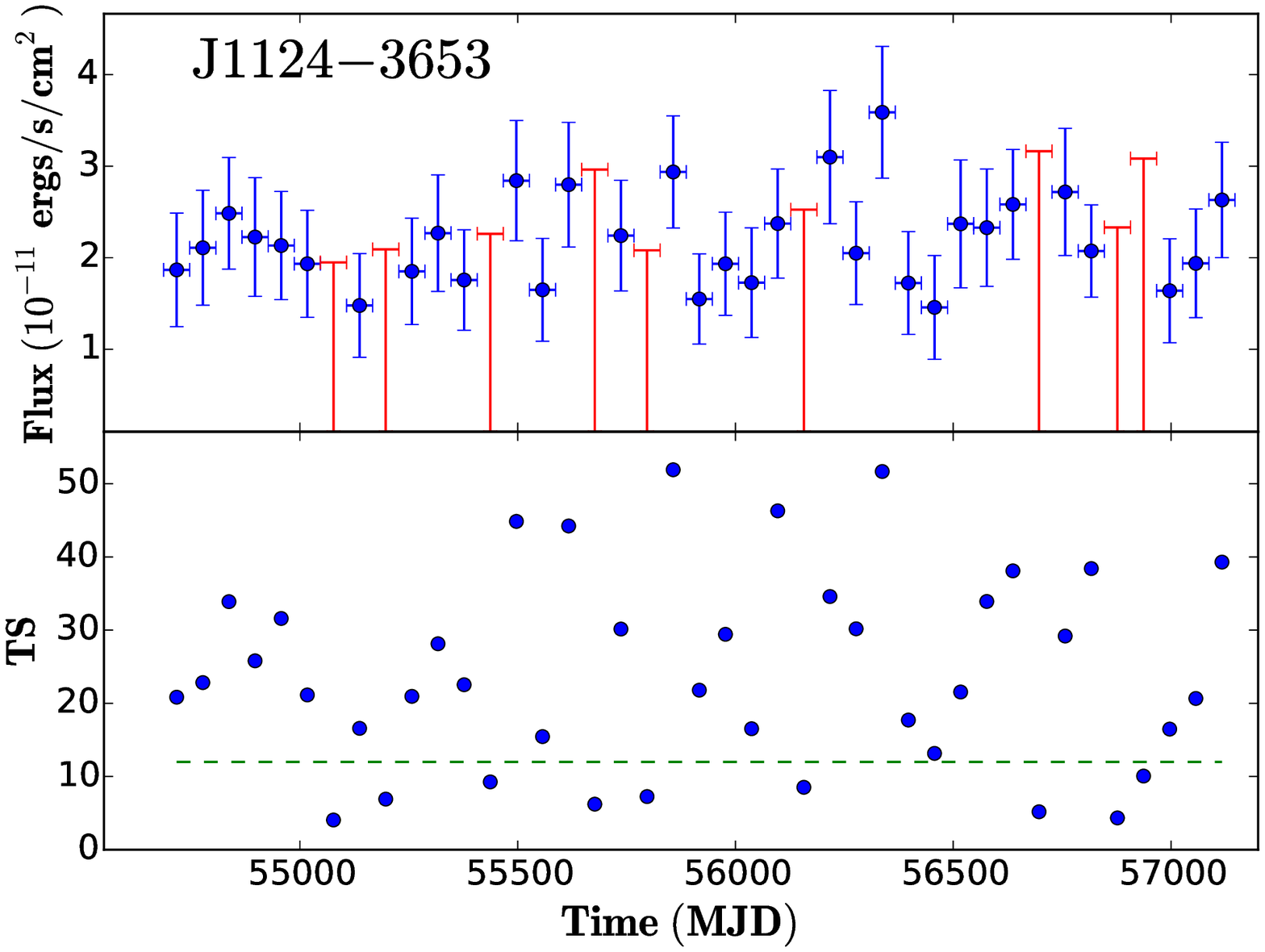}
\includegraphics[width=0.32\textwidth]{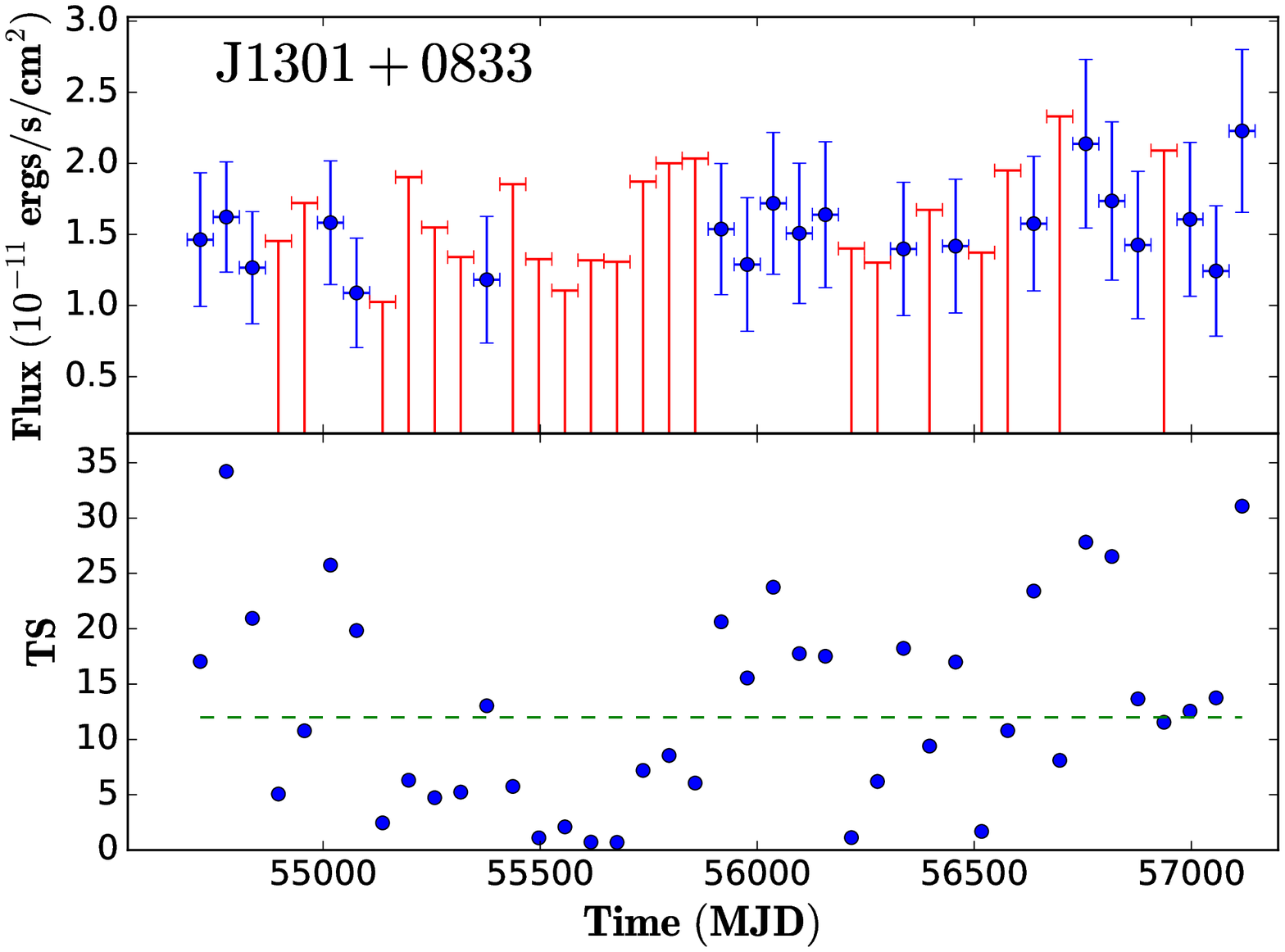}
\includegraphics[width=0.32\textwidth]{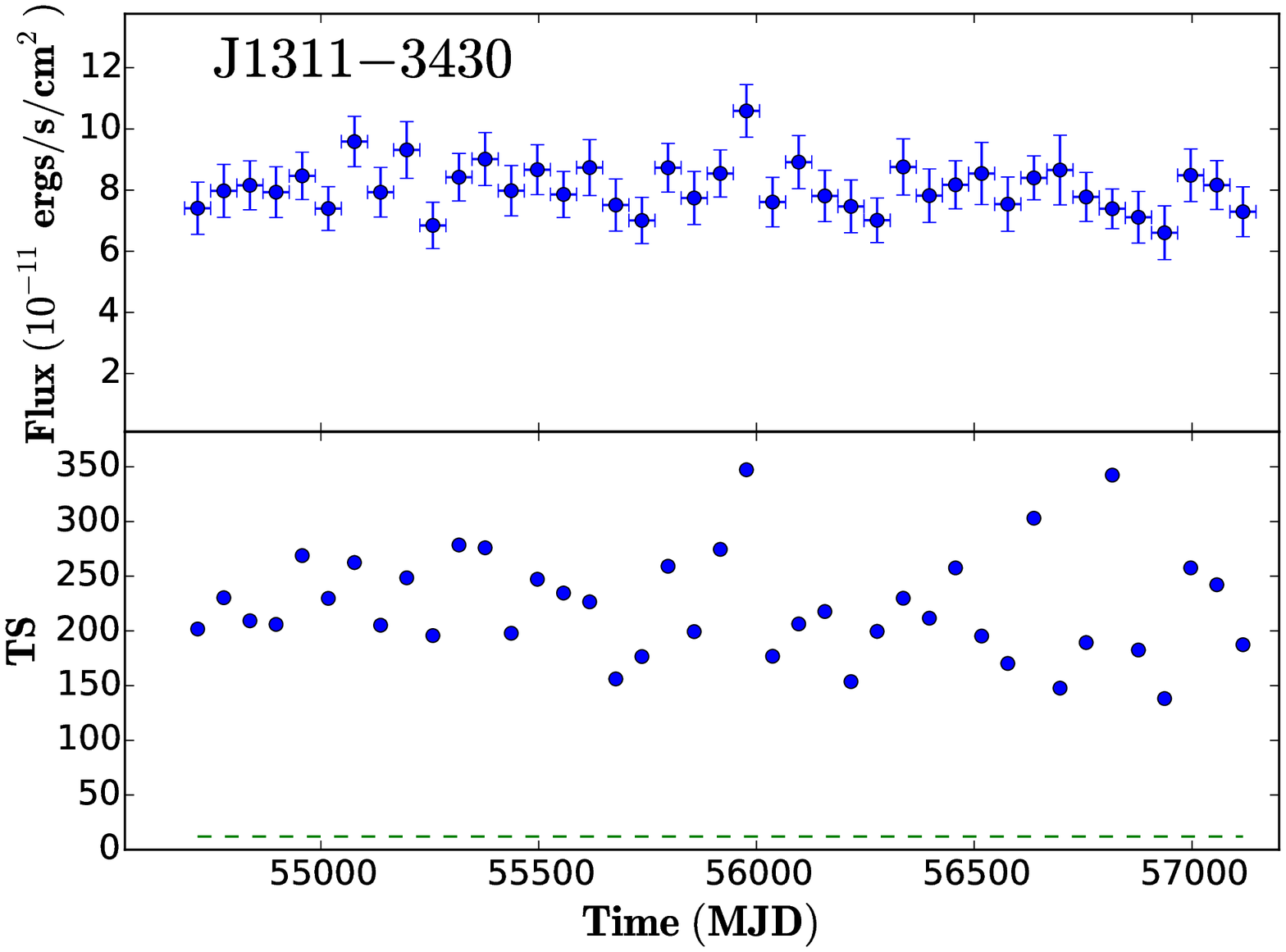}
\includegraphics[width=0.32\textwidth]{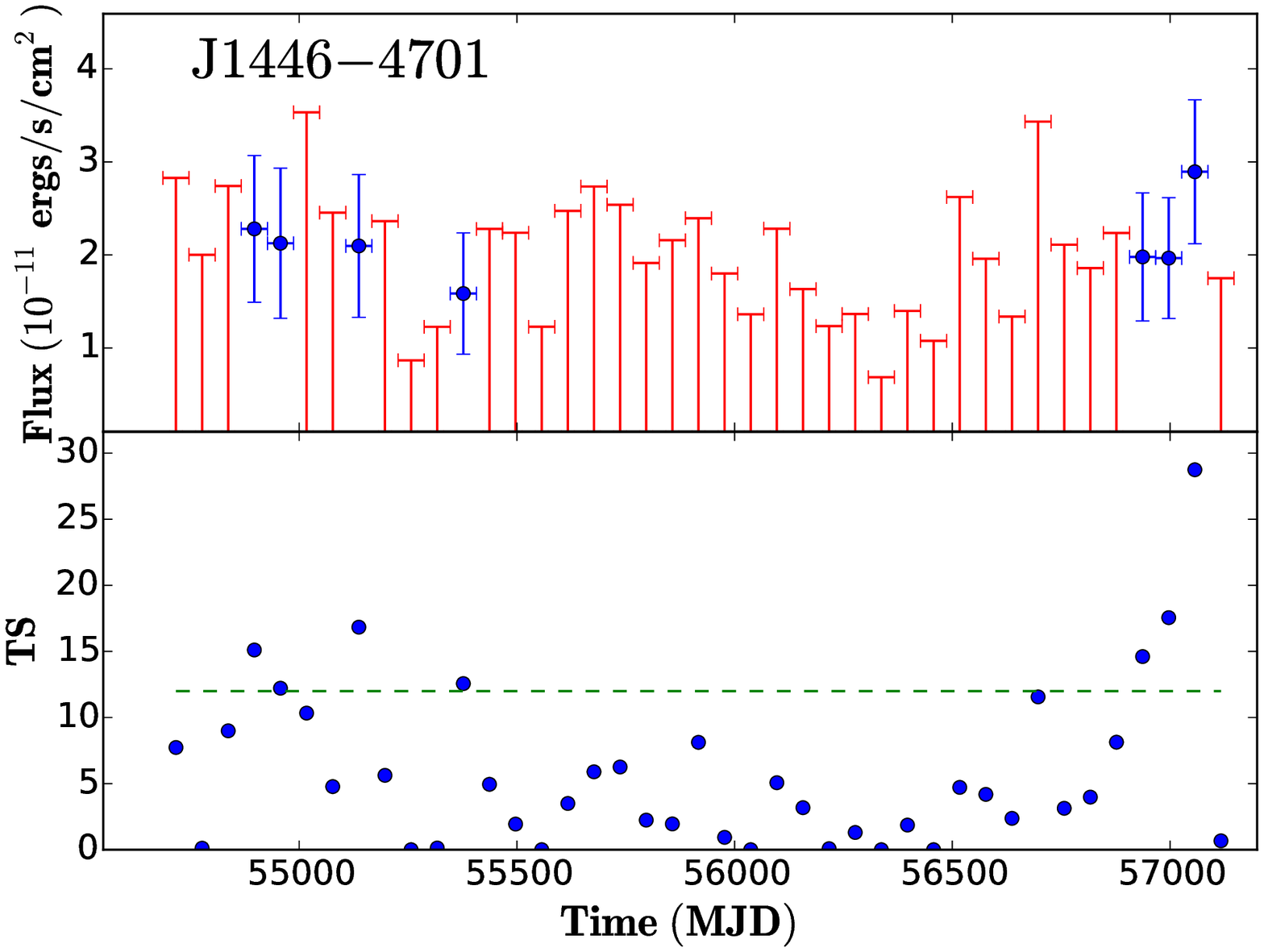}
\includegraphics[width=0.32\textwidth]{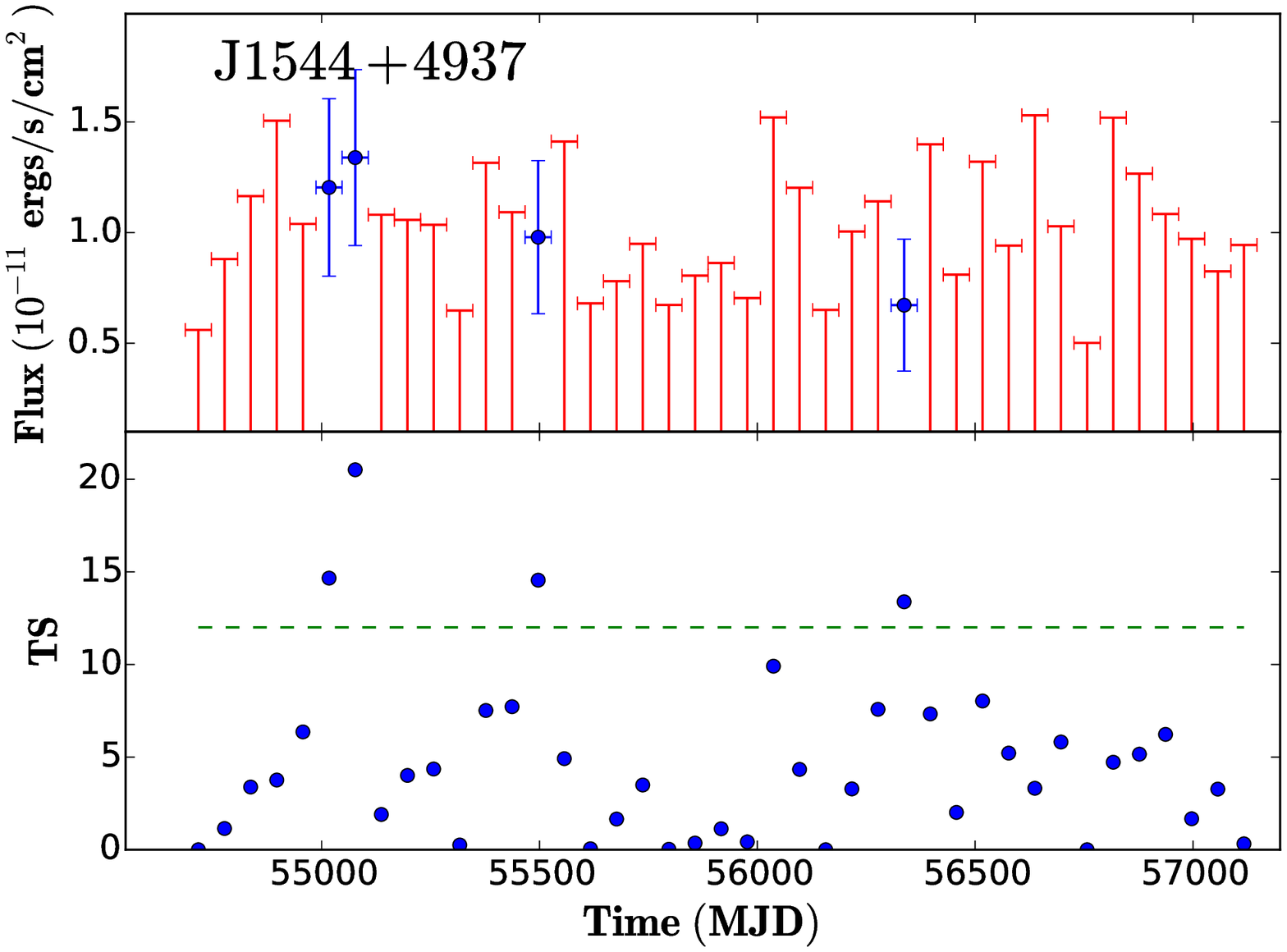}
\includegraphics[width=0.32\textwidth]{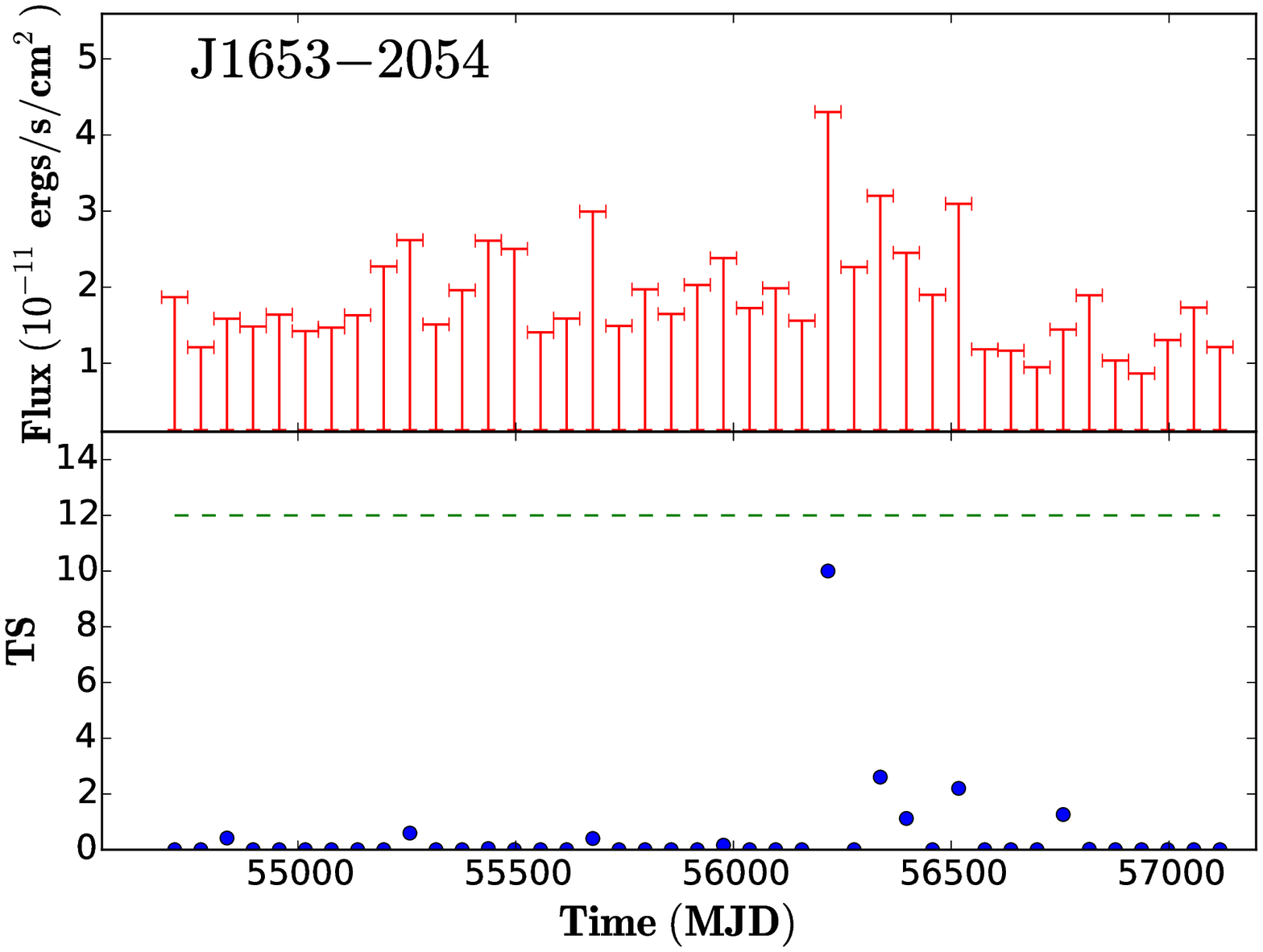}
\includegraphics[width=0.32\textwidth]{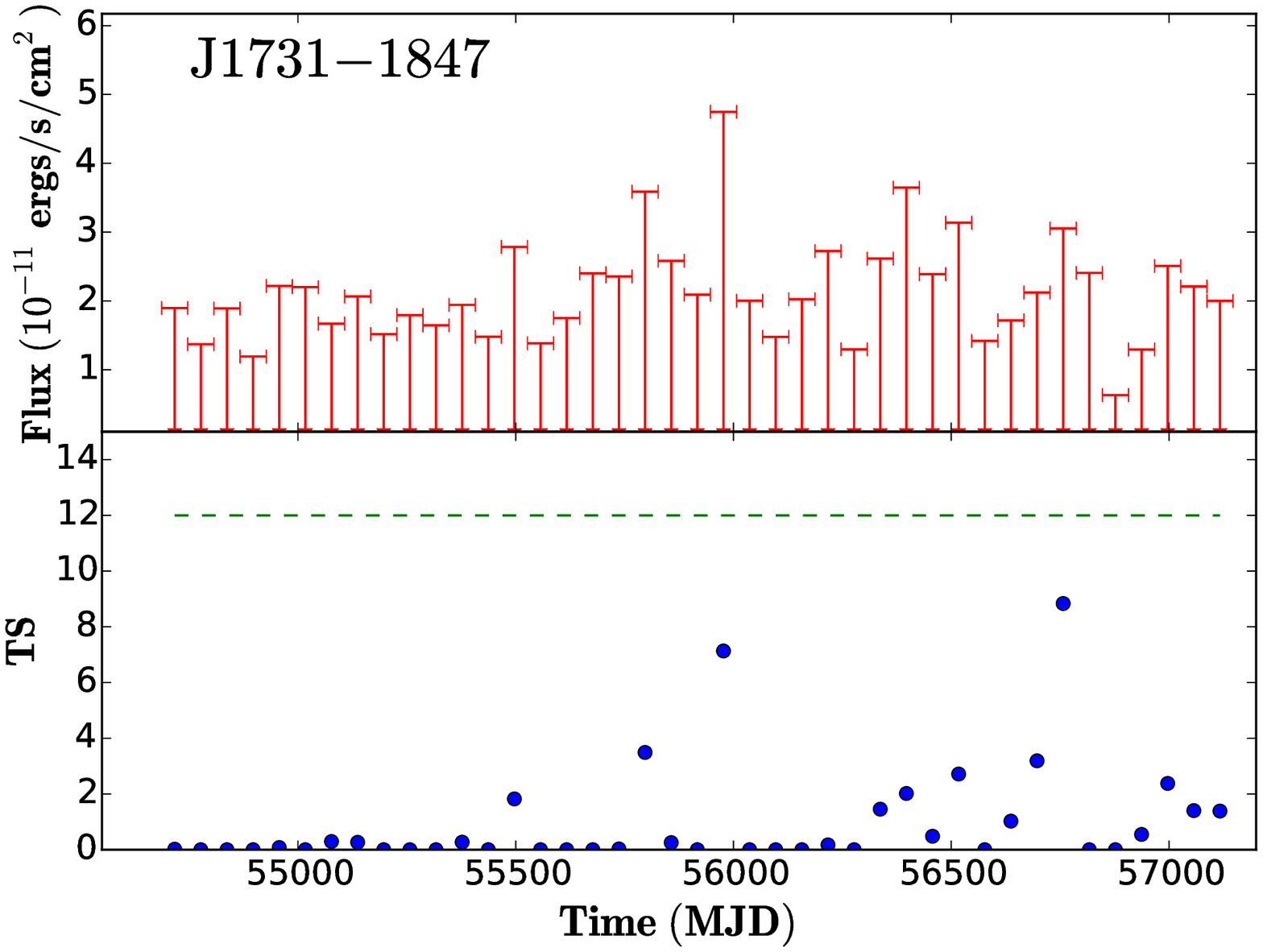}
\includegraphics[width=0.32\textwidth]{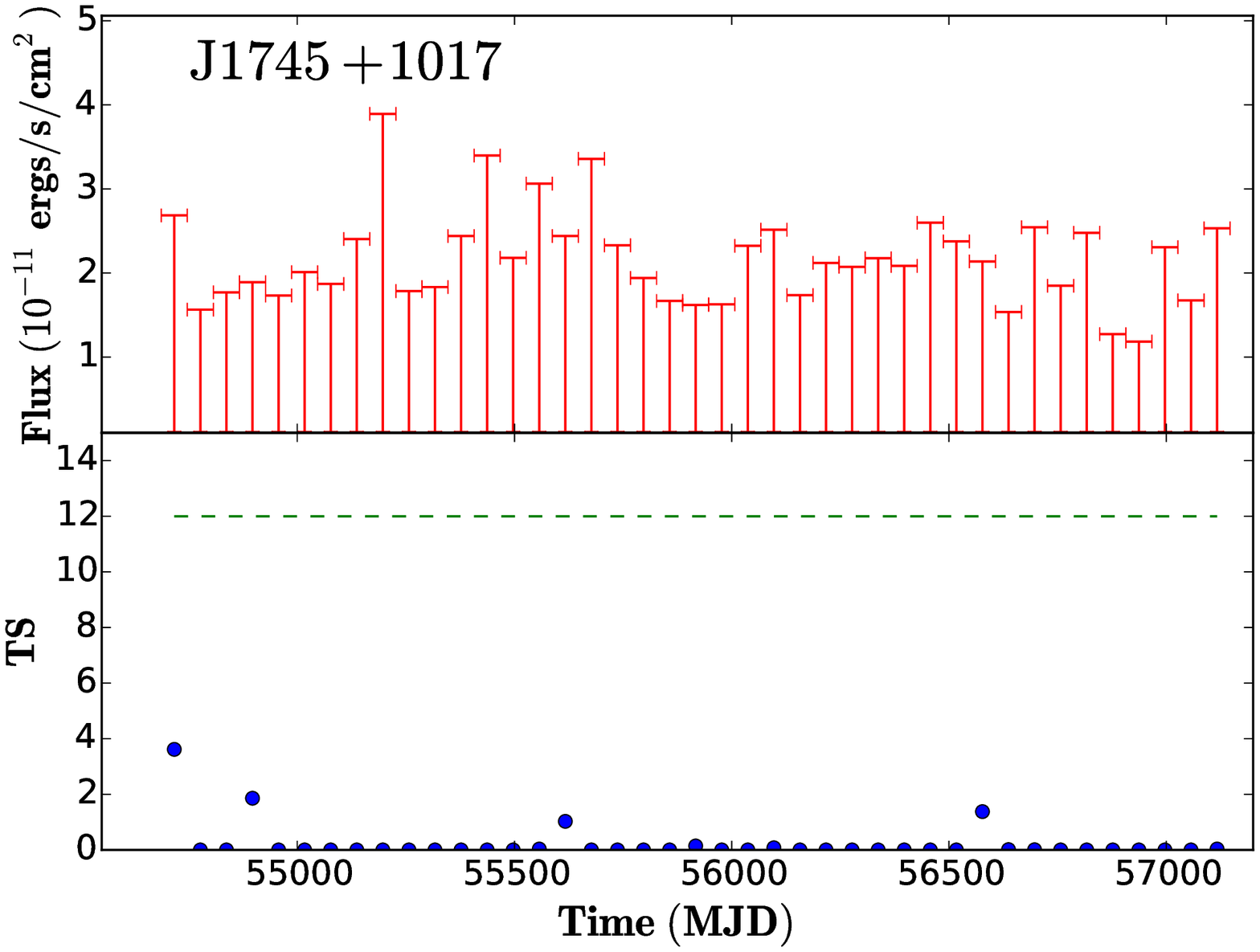}
\includegraphics[width=0.32\textwidth]{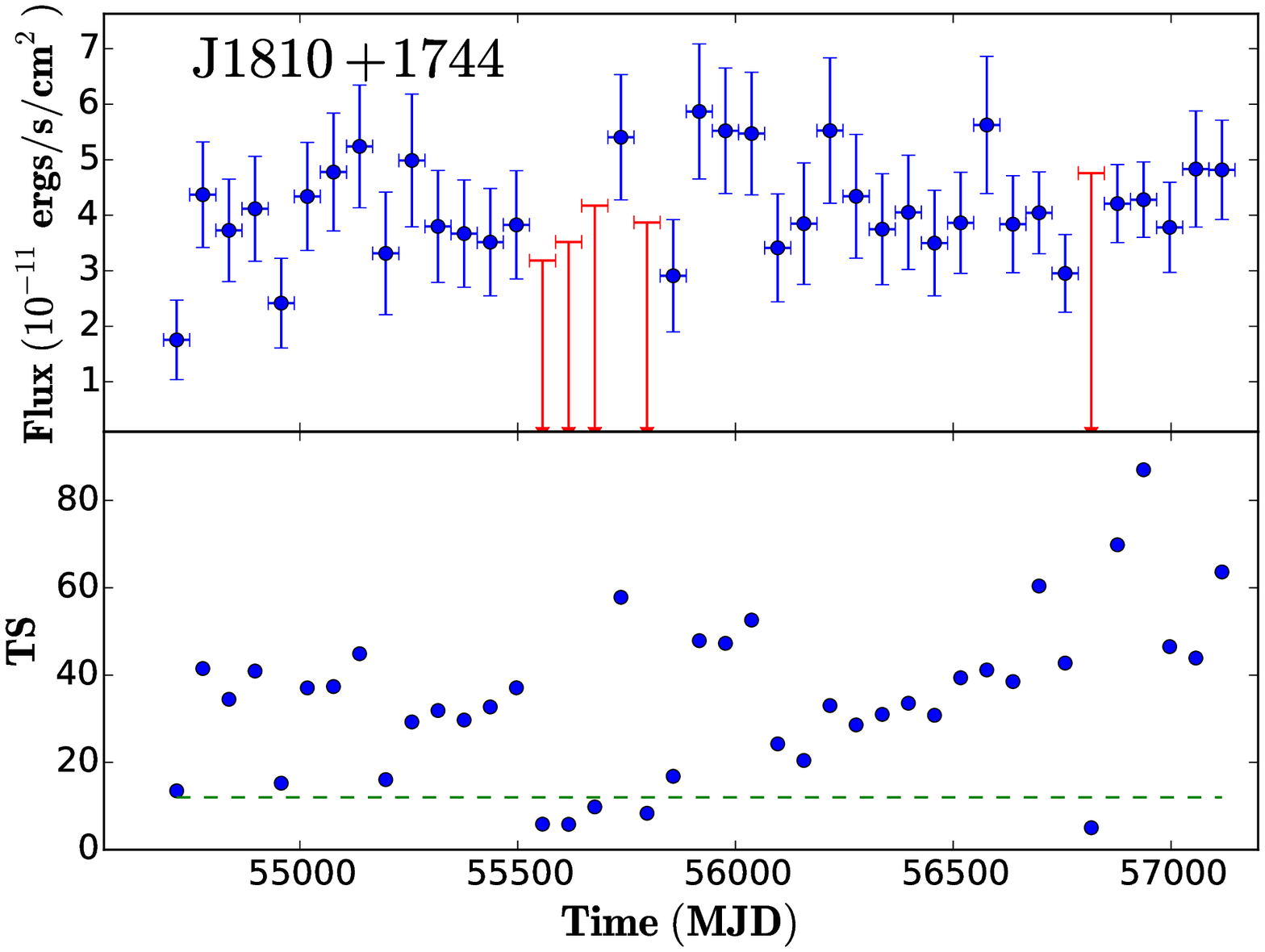}
\caption{{\it Continued} }
\end{figure}

\addtocounter{figure}{-1} \begin{figure}
\centering
\includegraphics[width=0.32\textwidth]{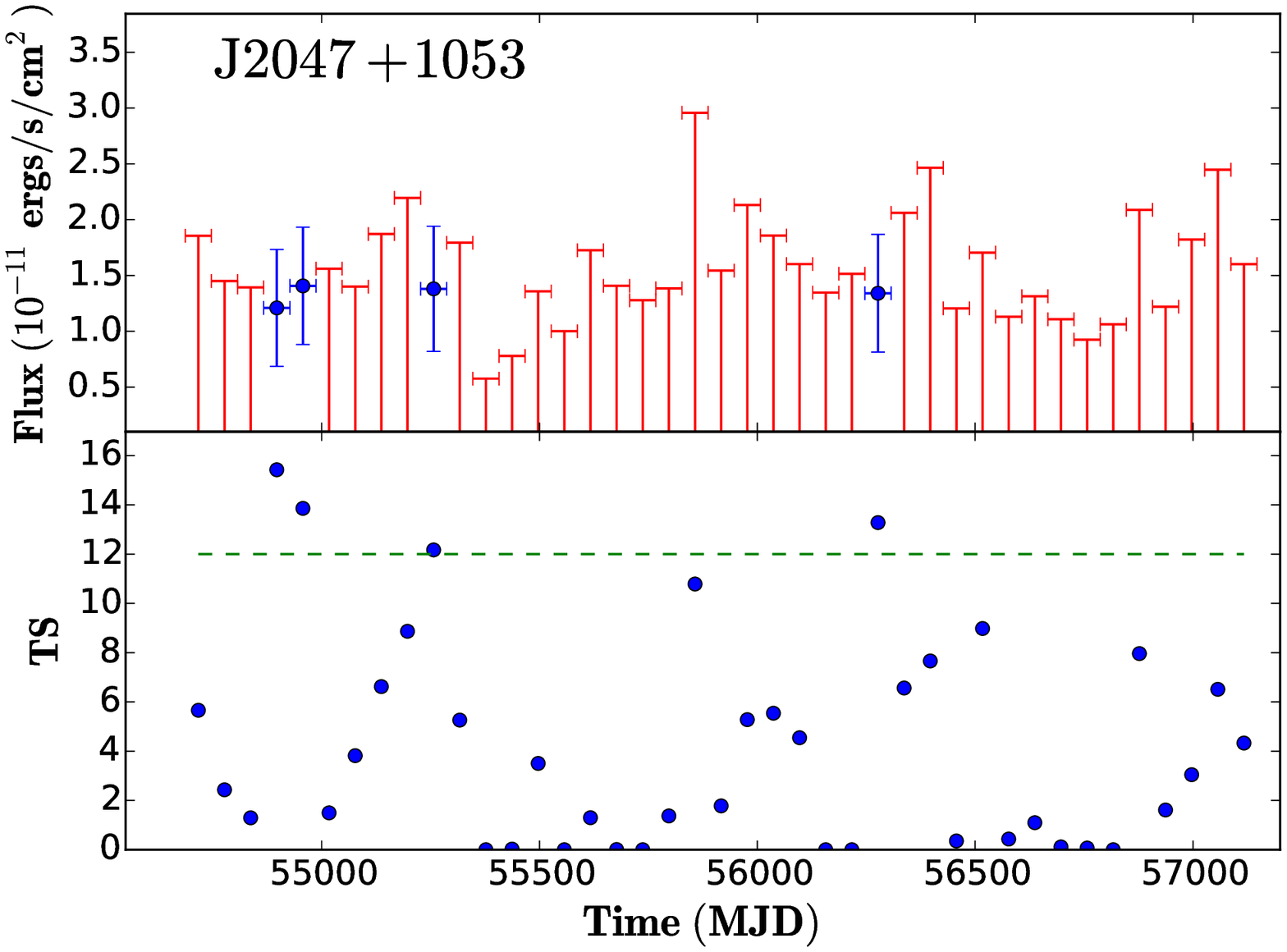}
\includegraphics[width=0.32\textwidth]{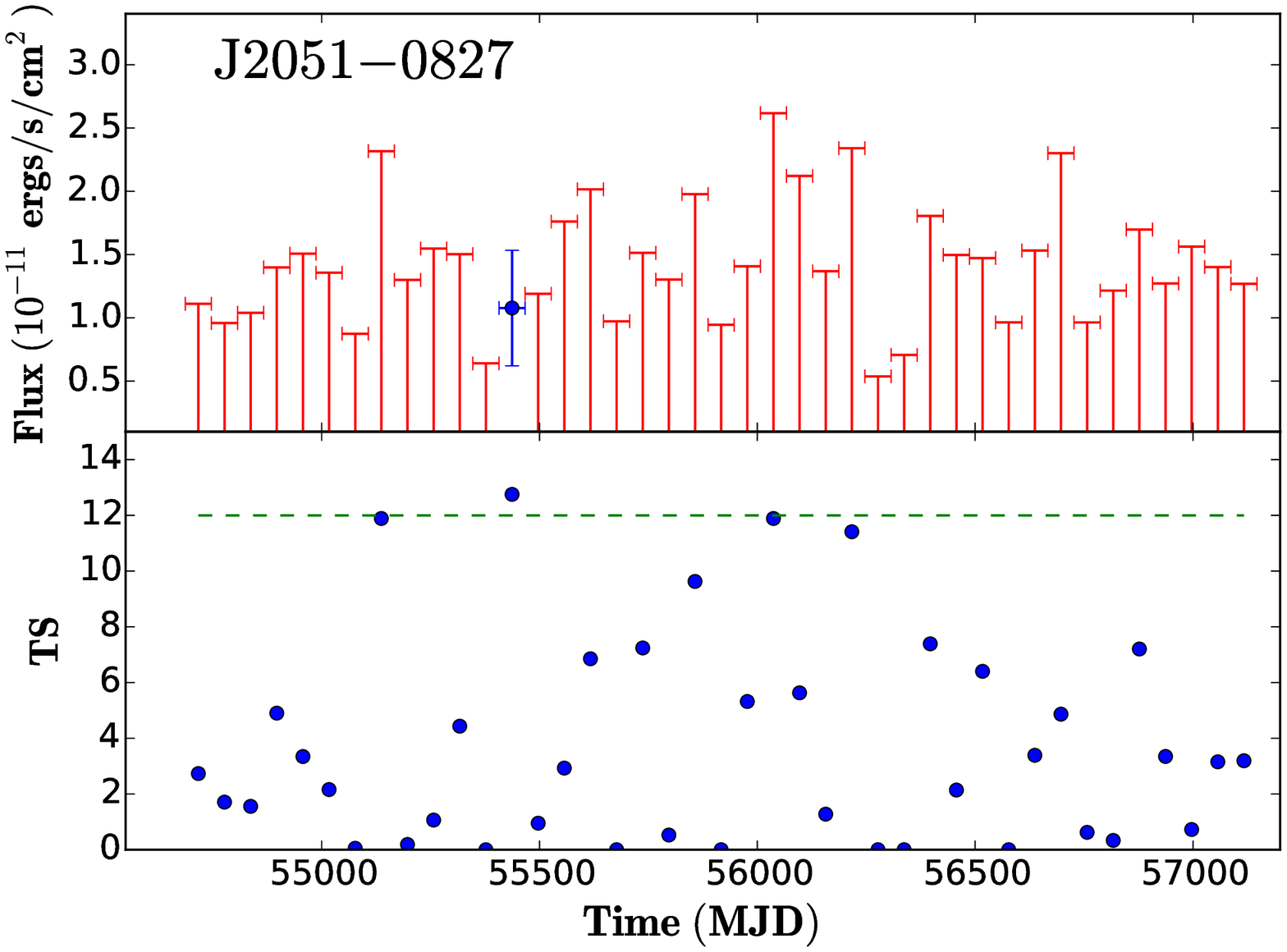}
\includegraphics[width=0.32\textwidth]{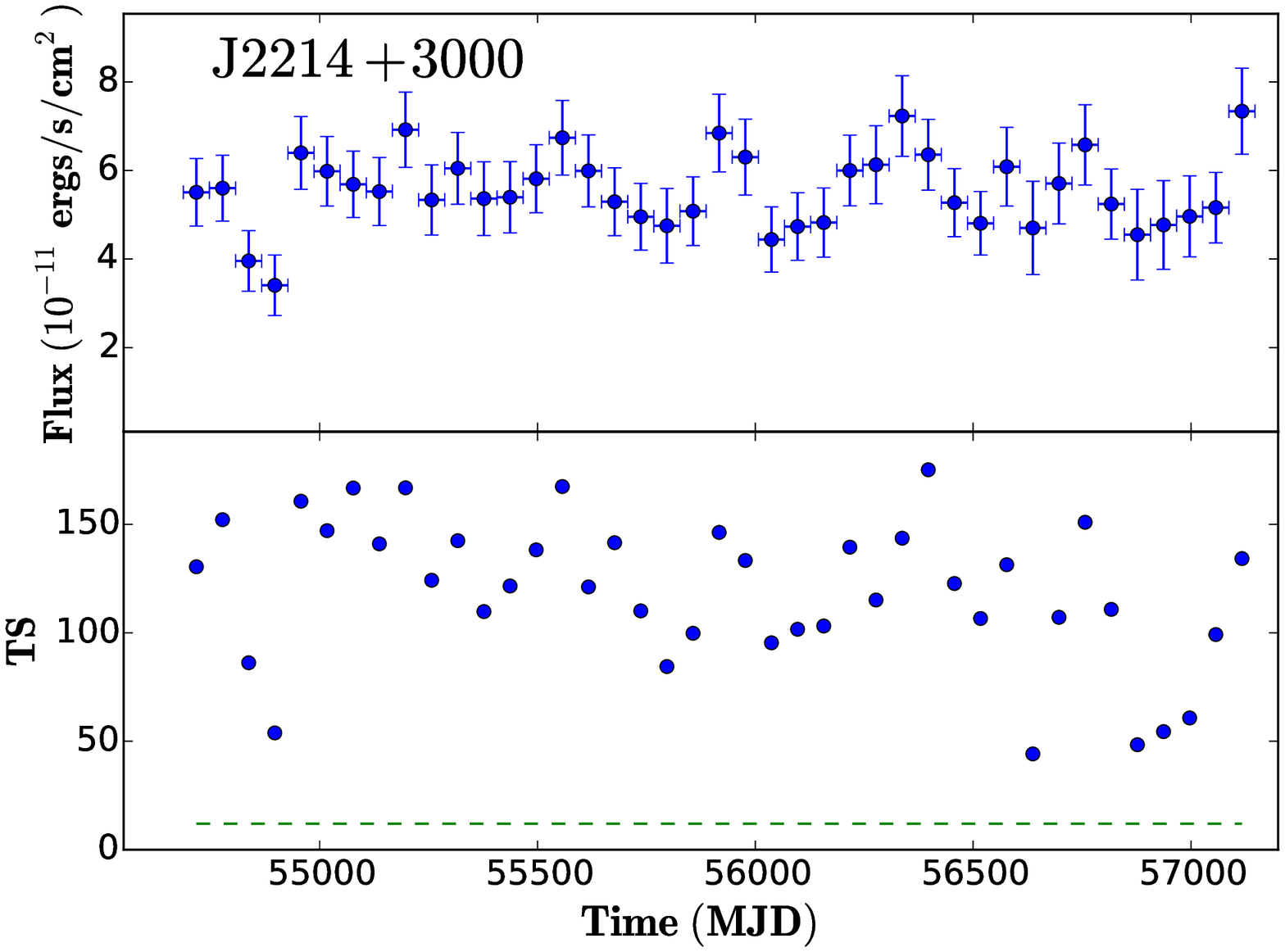}
\includegraphics[width=0.32\textwidth]{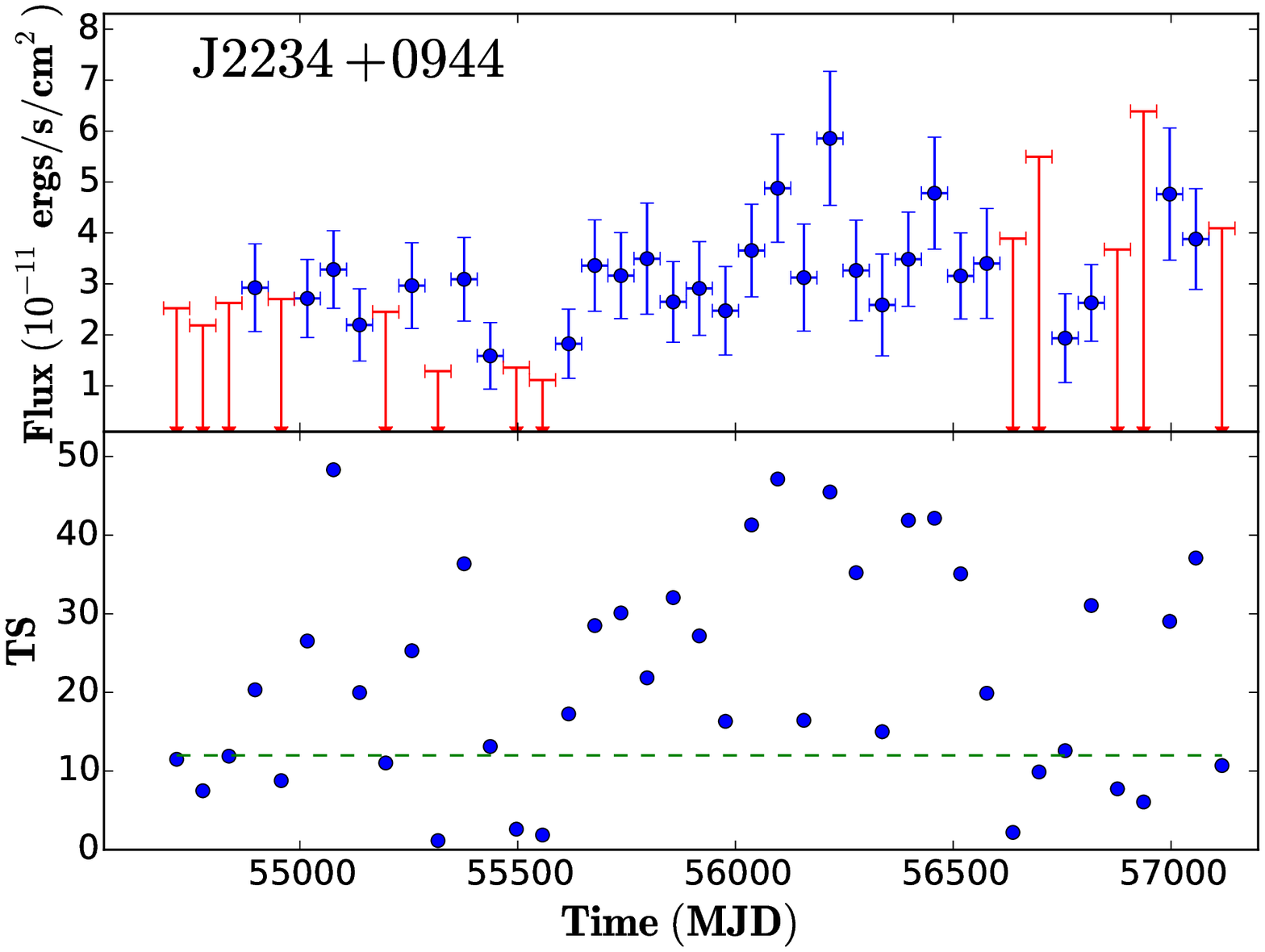}
\includegraphics[width=0.32\textwidth]{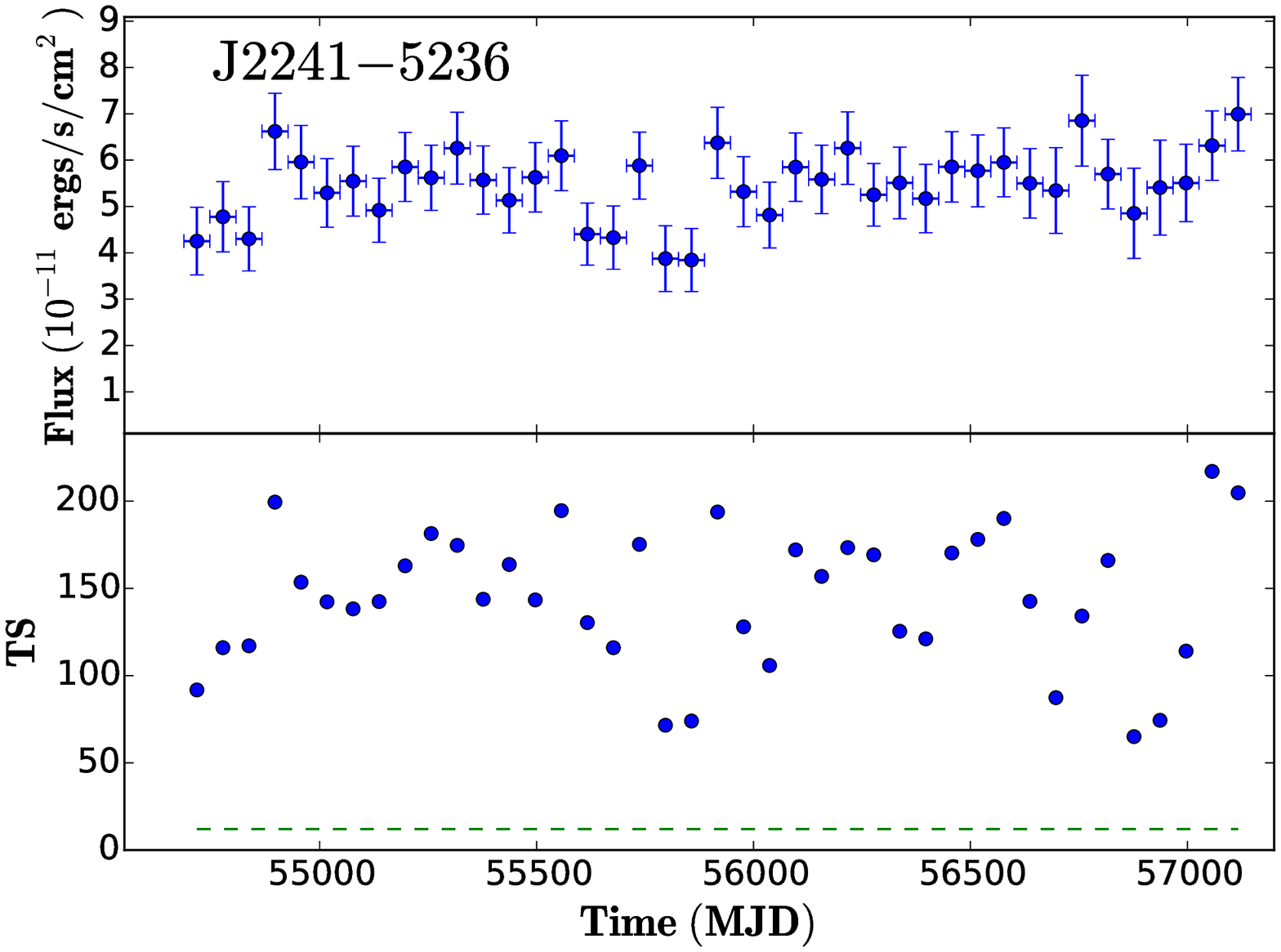}
\includegraphics[width=0.32\textwidth]{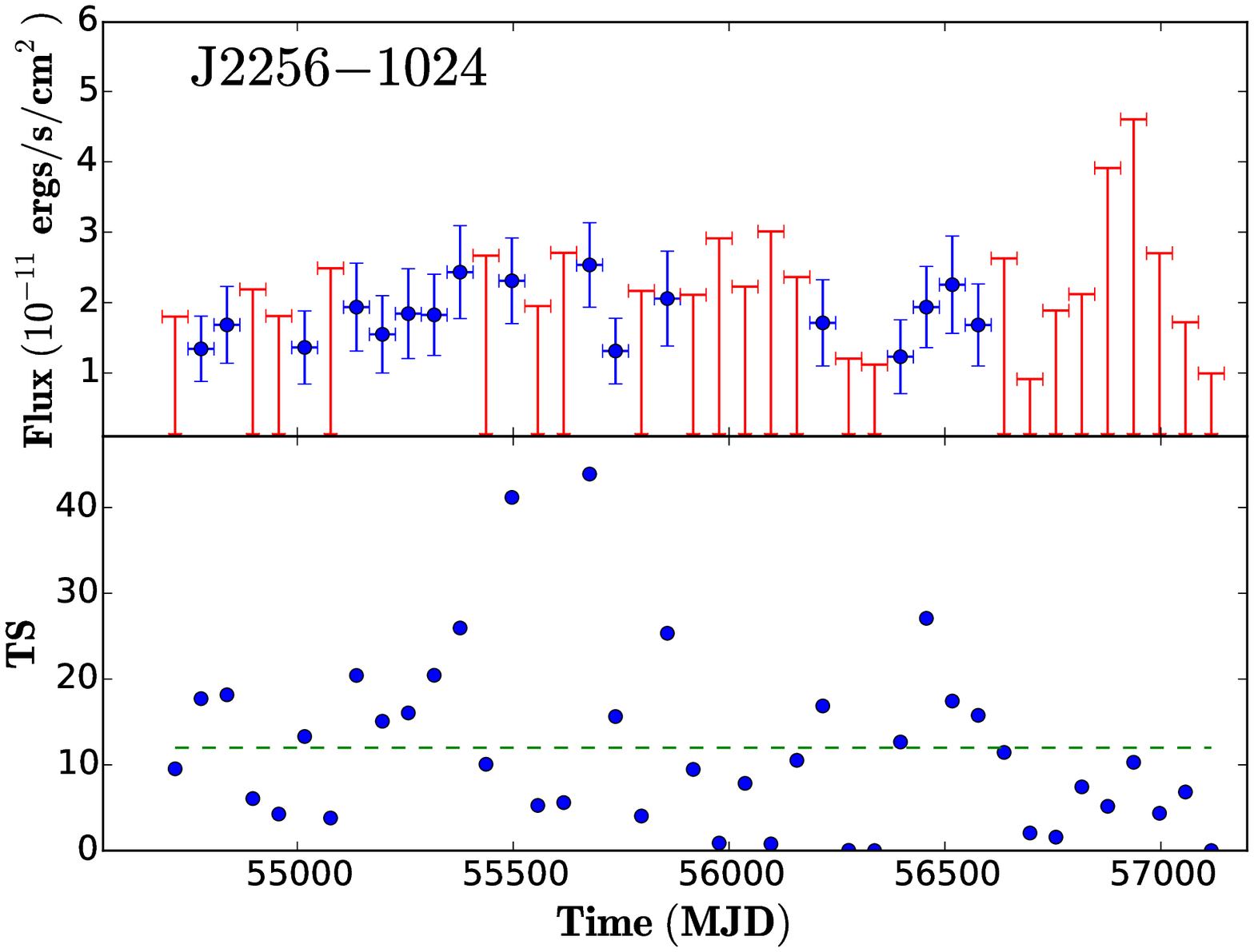}
\caption{{\it Continued} }
\end{figure}

%%%%%%%%%%%%%%%%%%%%%%%%%%%%%%%%%%%
\begin{center}
\begin{figure*}
\centering
\includegraphics[scale=0.32]{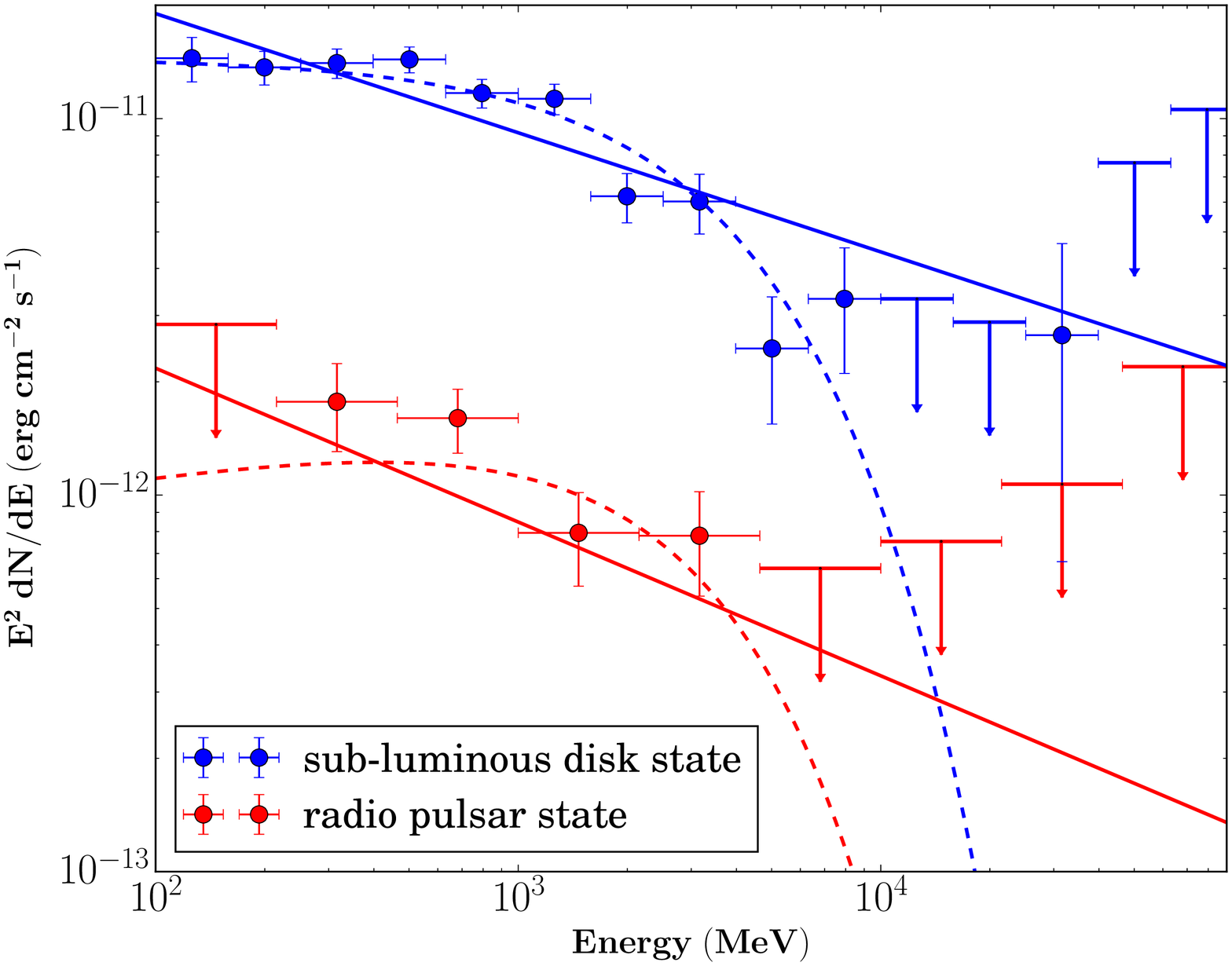}
\includegraphics[scale=0.32]{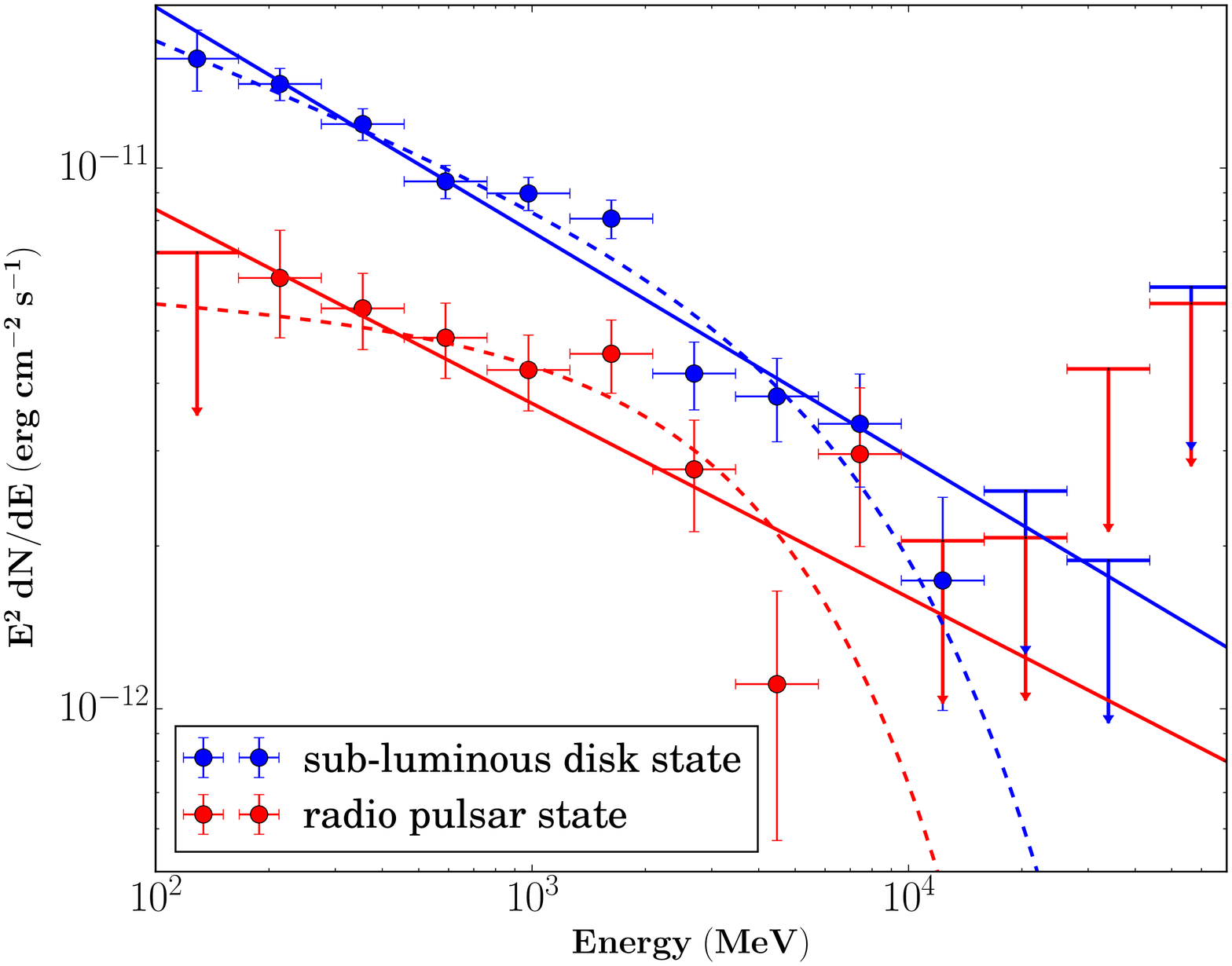}

\caption{\emph{Fermi}-LAT spectra of J1023+0038 (left) and J1227$-$4853 (right), in {radio pulsar state (red) and sub-luminous disk state (blue)}.
The corresponding \textit{gtlike}-fitted models are shown with solid lines (for a power-law) and dotted lines (for a {power law} with exponential cutoff).}
\label{SED}
\end{figure*}
\end{center}
%%%%%%%%%%%%%%%%%%%%%%%%%%%%%%%%%%

\begin{figure}
\centering
\includegraphics[width=0.4\textwidth]{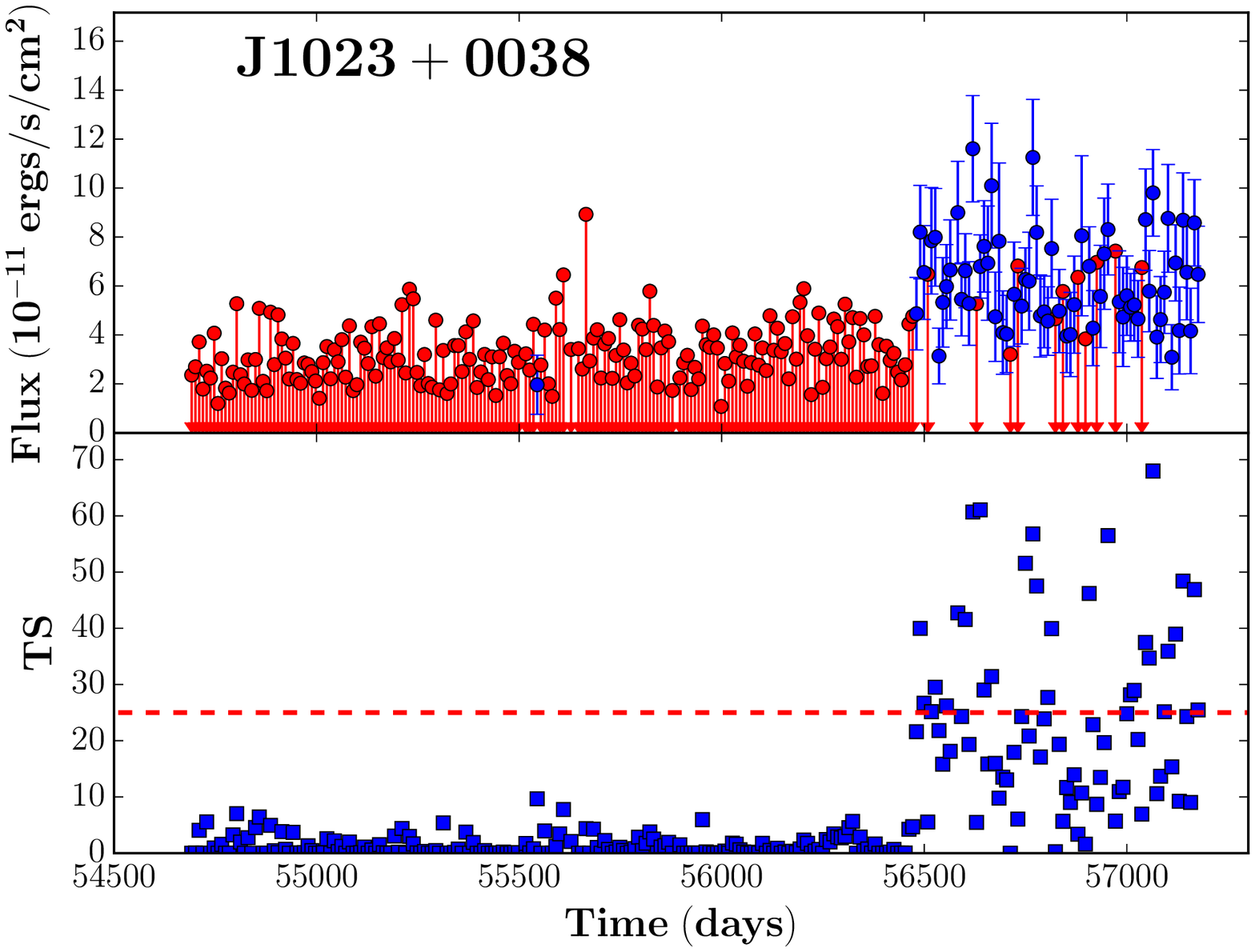}
\includegraphics[width=0.4\textwidth]{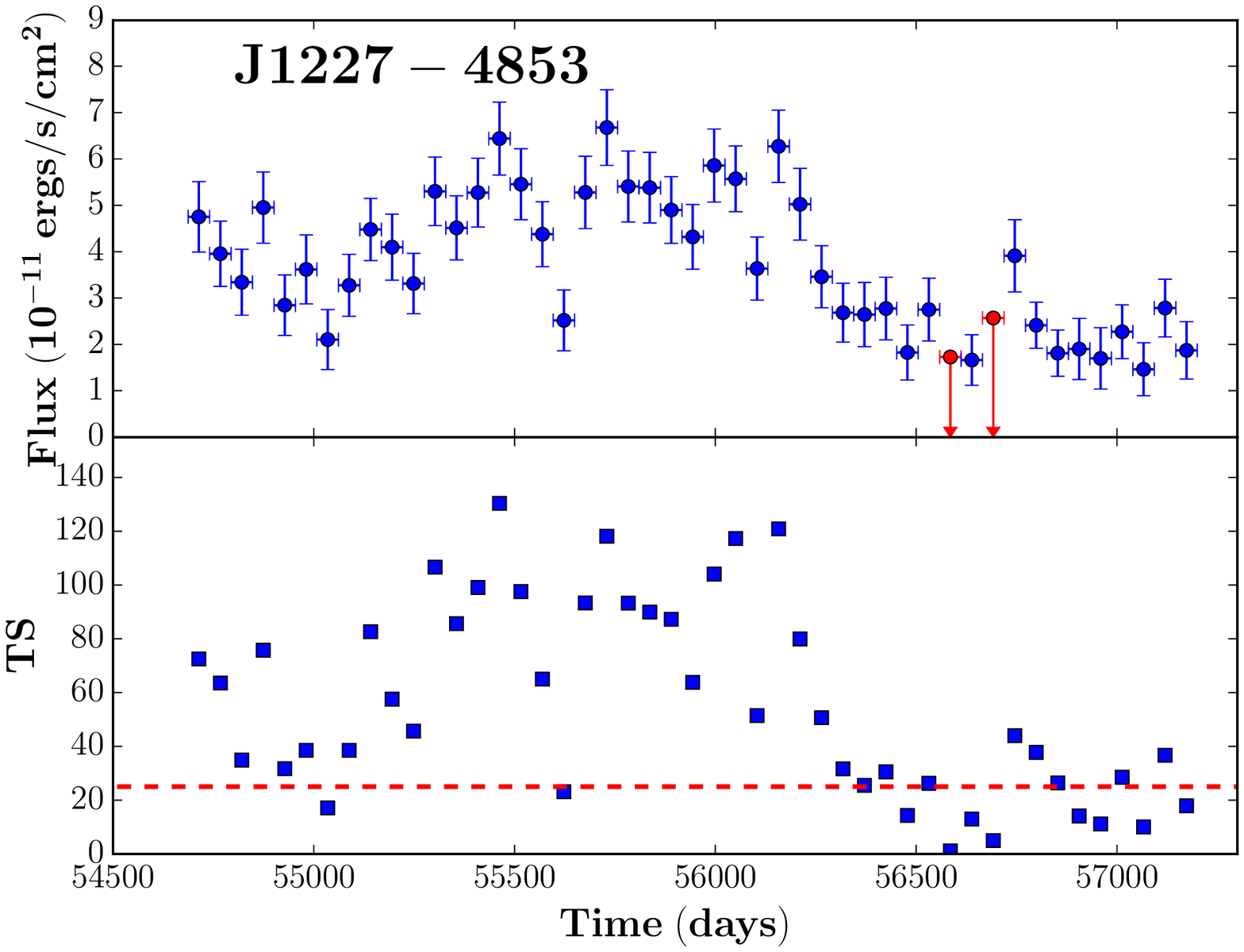}
\includegraphics[width=0.4\textwidth]{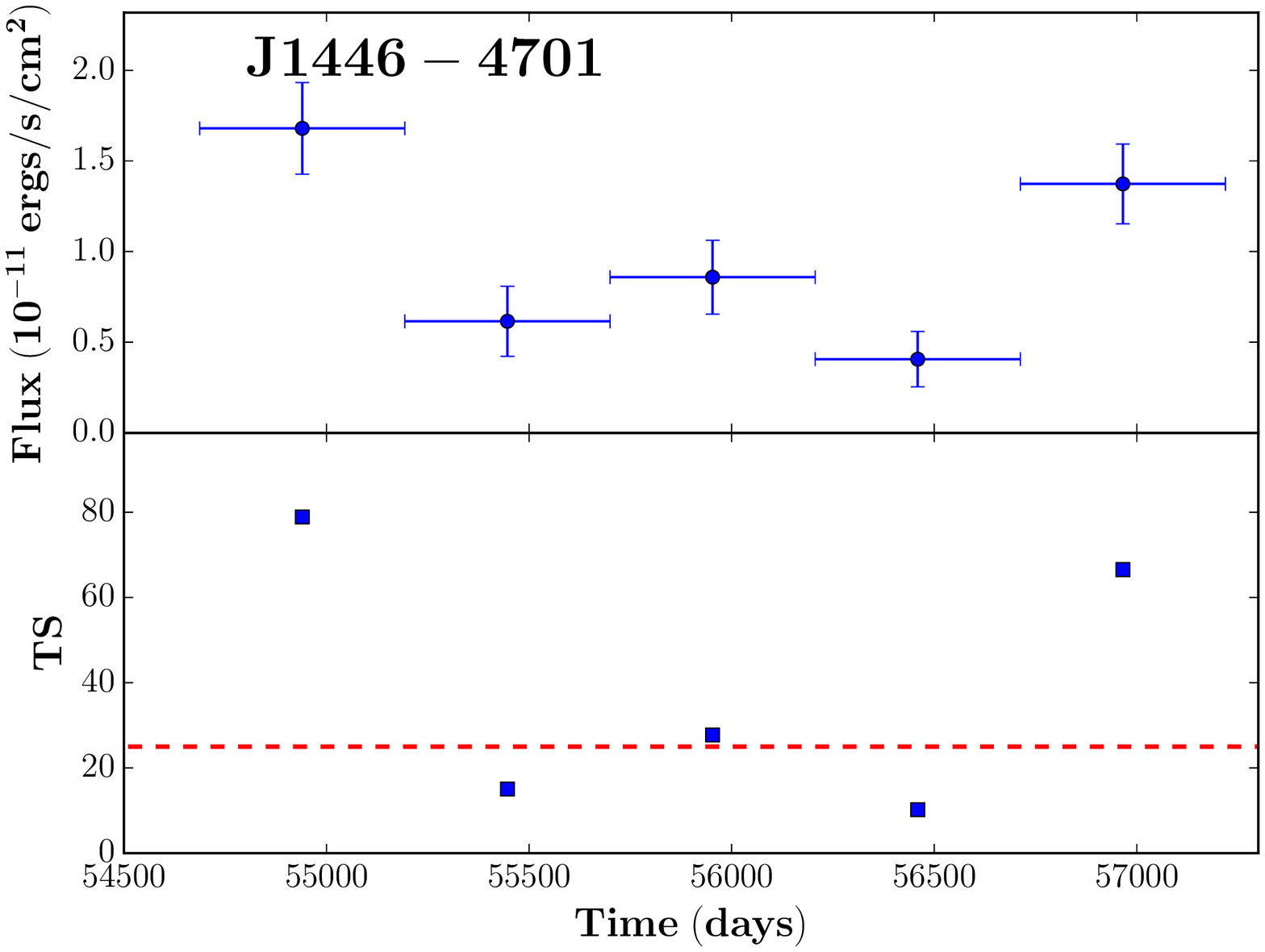}
\includegraphics[width=0.4\textwidth]{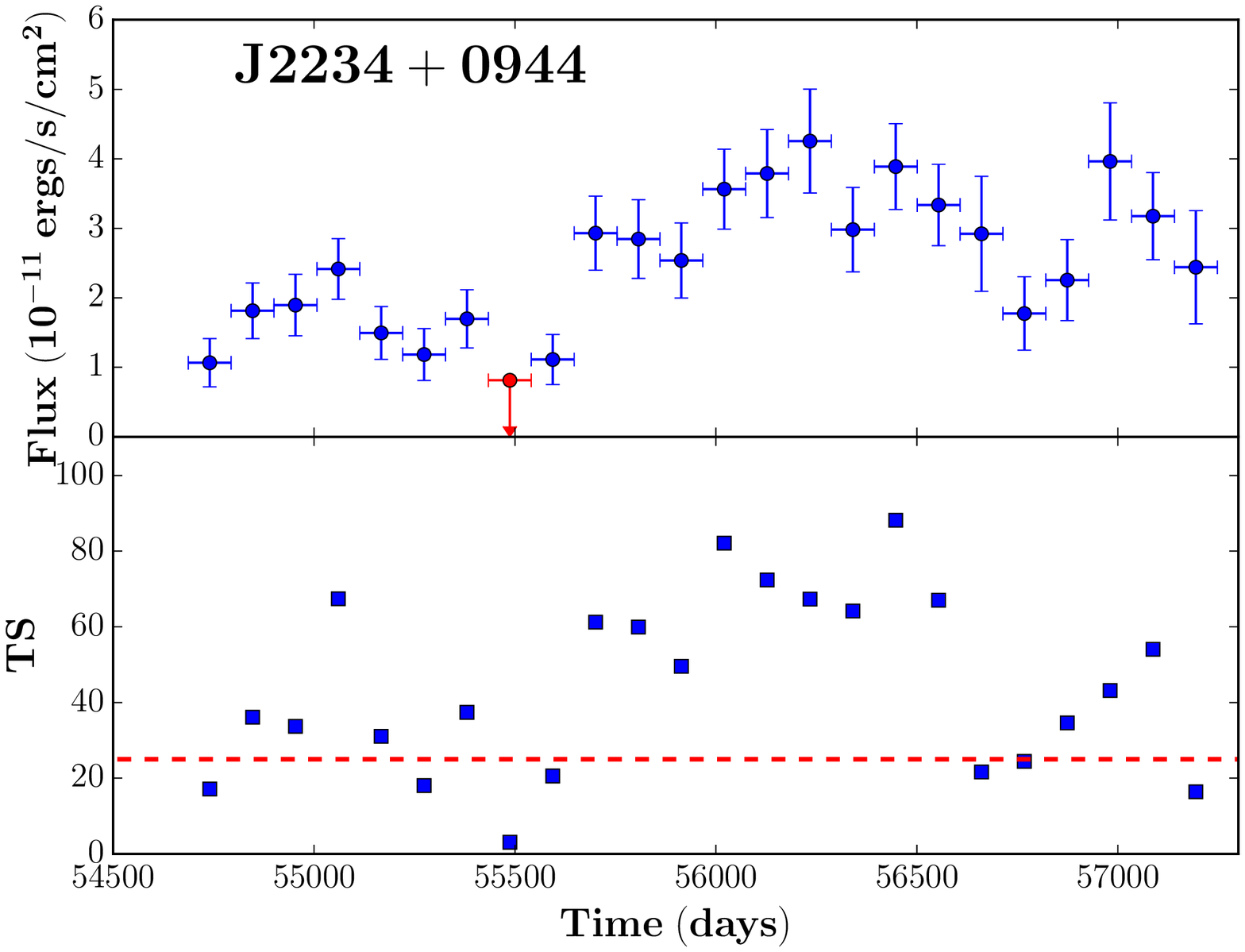}
\caption{Long-term light curves of {J1023+0038, J1227-4853,} J1446-4701 and J2234+0944, {for time bins} defined case-by-case via simulations.  The dotted horizontal line indicates TS=25. See text for details.}
\label{odd}
\end{figure}

\acknowledgments

The \emph{Fermi}-LAT Collaboration acknowledges support from a number of agencies and institutes for both development and the
operation of the LAT as well as scientific data analysis. These include NASA and DOE in the United States,
CEA/Irfu and IN2P3/CNRS in France, ASI and INFN in Italy, MEXT, KEK, and JAXA in Japan, and the K.~A.\ Wallenberg
Foundation, the Swedish Research Council and the National Space Board in Sweden. Additional support from INAF in Italy and CNES in
France for science analysis during the operations phase is also gratefully acknowledged.

We acknowledge the support from the grants AYA2015-71042-P, SGR 2014-1073 and the National Natural Science Foundation of
China via NSFC-11473027, NSFC-11503078, NSFC-11133002, NSFC-11103020, XTP project XDA 04060604
and the Strategic Priority Research Program ``The Emergence of Cosmological Structures" of the Chinese Academy of Sciences, Grant No. XDB09000000,
as well as the CERCA Programme of the Generalitat de Catalunya.
N.R. is further supported by an NWO Vidi Award.
A.P. acknowledge support via an EU Marie Sklodowska-Curie Individual Fellowship under contract No. 660657-TMSP-H2020-MSCA-IF-2014, as well as we all acknowledge fruitful discussion with the international team on ``The disk-magnetosphere interaction around transitional millisecond pulsars" at ISSI (International Space Science Institute), Bern.
We thank {T. J. Johnson and P. S. Ray } for comments.

\end{document}